\documentclass[aps,prb,reprint,groupedaddress,amsmath]{revtex4-1}

\usepackage{braket}
\usepackage{bm}
\usepackage{graphicx}
\usepackage{mathrsfs}
\usepackage{subfigure}
\usepackage{ulem}
\usepackage{hyperref}
\hypersetup{hypertex=true,
            colorlinks=true,
            linkcolor=blue,
            anchorcolor=blue,
            citecolor=blue}

\begin{document}


\title{Twist versus heterostrain control of optical properties of moir\'e exciton minibands}


\author{Huiyuan Zheng}

\author{Dawei Zhai}

\author{Wang Yao}

\affiliation{Department of Physics and Guangdong-Hong Kong Joint Laboratory of Quantum Matter, The University of Hong Kong, Hong Kong, China}
\affiliation{HKU-UCAS Joint Institute of Theoretical and Computational Physics at Hong Kong, China}


\date{\today}

\begin{abstract}
We investigate the optical properties of interlayer excitons in heterobilayer transition metal dichalcogenides where moir\'e pattern is introduced by heterostrain, in comparison with that introduced by twisting (and/or lattice mismatch). Besides being a cause of the moir\'e texture, strain also effectively introduces a constant gauge potential on either electron or hole, which shifts the dispersion of kinetic energy with respect to the excitonic crystal momenta in the moir\'e superlattices. This leads to distinct exciton mini-band dispersions and light coupling properties from the twisting induced moir\'e, even if the excitonic moir\'e superlattice potentials have the similar real-space profile for the two cases. For strain that breaks the three-fold rotational symmetry at the atomic scale, the exciton wave packets trapped at the superlattice potential minima have elliptically polarized valley optical selection rules, in contrast to the circularly polarized ones in the twisting moir\'e. We investigate the evolution of the excitonic mini-bands and the optical dipoles of the bright states inside the light cones with the decrease of the moir\'e periodicity, upon which the excitonic wave functions evolve from localized wave packets to the extended Bloch states. Furthermore, moir\'e exciton properties under the interplay of twisting and heterostrain are also discussed.
\end{abstract}


\maketitle


\section{Introduction\label{section_intro}}
Atomically thin group-VIB transition metal dichalcogenides (TMDs) have emerged as a class of two-dimensional (2D) semiconductors with exciting optical properties. The monolayers feature direct band gaps in the visible frequency range \cite{mak2010atomically,splendiani2010emerging}, with band edges located at the degenerate $\mathbf{K}$ and $-\mathbf{K}$ corners of the hexagonal Brillouin zone (BZ), which constitute the valley degree of freedom. Optical properties of TMDs are dominated by the hydrogen-like bound states of the electron and hole at the valleys, where strong Coulomb interaction leads to exceptionally large exciton binding energies \cite{yu2015valley}. As the monolayers are only bounded by weak van der Waals (vdW) interactions, heterostructures of different TMDs can be flexibly engineered without the requirement of lattice matching, which allows vast opportunities to engineer and extend optoelectronic properties. TMD heterobilayers typically feature the type-II alignment where electron and hole energetically favor the two opposite layers. Interlayer excitons (IXs) with the layer separation of the electron and hole ingredients therefore becomes the lowest energy configuration that can dominate the photoluminescence \cite{rivera2015observation}. Compared with monolayer excitons, IXs in TMD bilayers exhibit ultralong recombination lifetime and spin-valley lifetime due to the reduced electron-hole wave function overlap, and electrically tunable resonance and strong dipolar interaction due to the permanent electrical dipole, providing rich possibilities in optoelectronic applications \cite{gong2014vertical,ceballos2014ultrafast,fang2014strong,rivera2015observation,rivera2016valley,lin2015atomically}.

Without the requirement of lattice matching, small differences in lattice constant and crystalline orientation lead to the formation of moir\'e patterns, i.e. spatial modulation on atomic registries from the interference of two mismatched atomic lattices. The sensitive dependence of electronic structures on the atomic stacking registries therefore results in the spatial variation in the local band gap in heterobilayers \cite{ zhang2017interlayer}. The  moir\'e-patterned band gap corresponds to the spatial modulation of IX energy, which can be tuned electrically by means of Stark effect, while the spatial profile can be engineered through the variation of twisting angles~\cite{wang2012three}. IXs in such moiré potential can be exploited in two functioning regimes: (i) excitonic emitters localized in trap arrays in relatively large moir\'e; and (ii) excitonic superlattices in relatively small moir\'e, where exciton hopping leads to formation of mini-bands \cite{yu2017moire}. Evidences of moir\'e excitons in both regimes are reported with distinct spectroscopic features in the various heterobilayers \cite{seyler2019signatures,tran2019evidence,jin2019observation,alexeev2019resonantly, brotons2020spin, li2020dipolar}. However, a systematic analysis of how the moir\'e exciton optical properties vary between the two regimes is still lacking, which is one of the issues that we address in this work.

\begin{figure*}
	\centering
	\setlength{\abovecaptionskip}{0.cm}
	\includegraphics[width=1\textwidth]{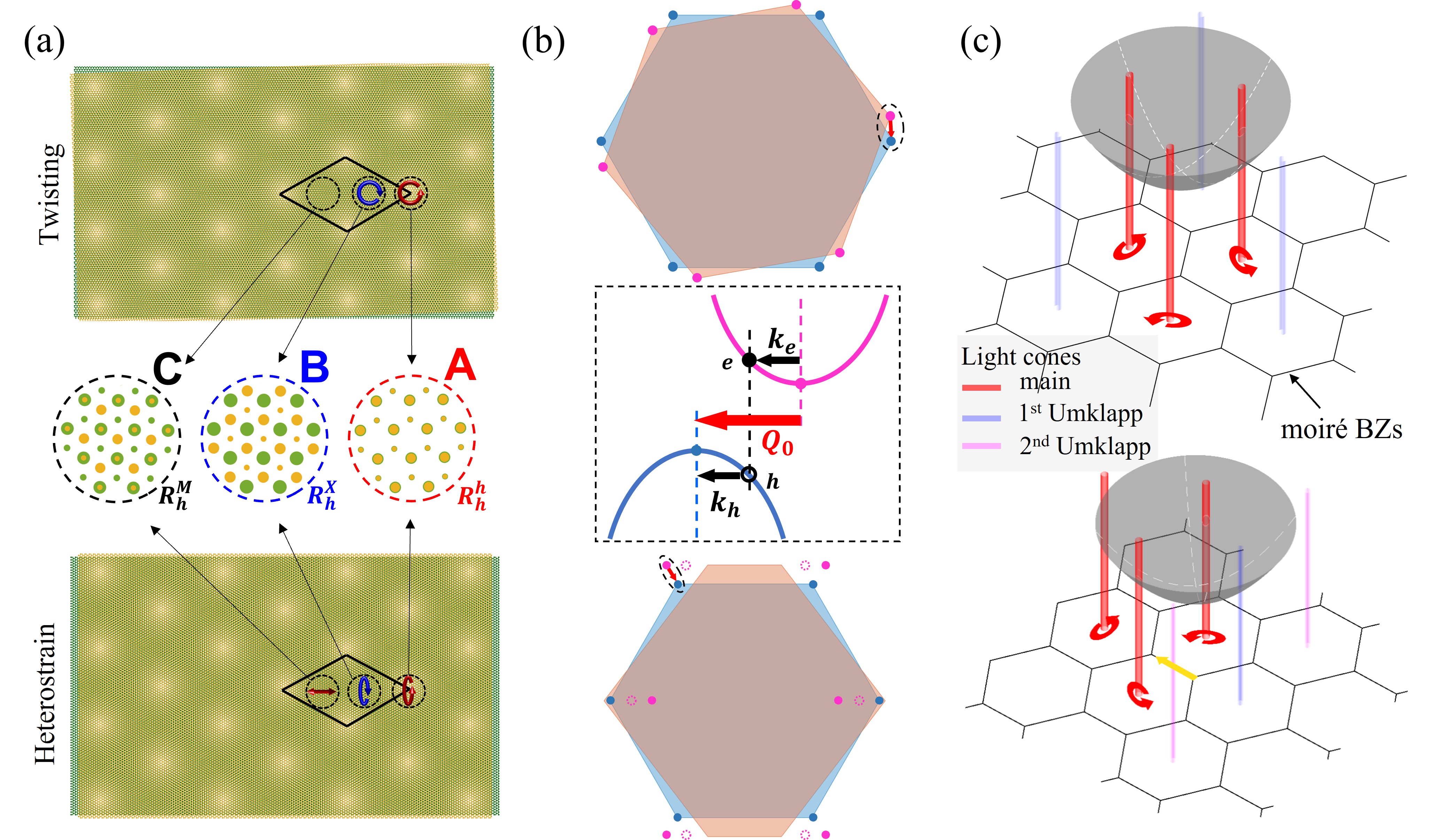}
	
	\caption{Comparison between interlayer excitons in twisting and heterostrain induced moir\'e patterns. (a) Real space configuration of bilayers subject to twisting (upper panel) or volume-preserving heterostrain (lower panel). Black rhombus depicts a moir\'e supercell. The in-plane polarizations of a $\mathbf{K}$ valley exciton at different high symmetry locations (dashed circles) are schematically shown. At $R^h_h$ and $R^X_h$ stacking locals, circular (elliptical) polarization is allowed in the twisted (heterostrained) moir\'e. At the $R^M_h$ stacking local in-plane linear polarization is forbidden (allowed) in twisted (heterostrained) moir\'e. (b) Mismatched monolayer BZs of twisted and heterostrained bilayers. The blue and red color represent the bottom and top layer, respectively. Blue (magenta) solid dots are Dirac points of the bottom (top) layer. For the strain case, we show both the Dirac point locations with (solid dots) and without (empty dots) considering the strain induced gauge potential. Red arrows mark the center-of-mass kinematic momentum of a bright IX, which correspond to the mismatch of the Dirac points in the two layers. The middle dashed box illustrates the various momenta in a bright IX with electron and hole from each layer. (c) Distinct distribution of light cones (cylinders) with respect to the IX kinetic energy dispersion (gray parabola) in the twisted (upper panel) and heterostrained (lower panel) moir\'e. Compared to twisted moir\'e, two major differences are noted in heterostrained moir\'e: (1) Exchange of the elliptical polarization at the three main light cones; (2) Overall shift of the light cone locations (yellow arrow) with respect to the kinetic energy dispersion.
	}
	\label{fig_twistvsstrain}
\end{figure*}

On the other hand, moir\'e patterns can also be created by applying layer-dependent strain (i.e. heterostrain) to lattice matched bilayers \cite{zhang2019magnetotransport,yu2020giant,zhai2020theory}. In particular, under a volume preserving strain, the bilayer moir\'e will exhibit a large-scale interference pattern nearly indistinguishable from that introduced by twisting and/or lattice mismatch (Fig.~\ref{fig_twistvsstrain}a). Compared to twisting, the heterostrain approach allows in situ tunability of the moir\'e, where the periodicity can in principle be controlled via substrates mechanically, thermally or piezoelectrically~\cite{frisenda2017biaxial,roldan2015strain,deng2018strain,yang2021strain,han2021experimental}. Besides, strain can be unintentionally introduced due to unavoidable deformation in fabrications of moir\'e superlattices, where the interplay of uniaxial strain and twisting can lead to dramatic elongation of the moir\'e towards one-dimensional structure. Linearly polarized IX photoluminescence in the strain elongated moir\'e traps has been reported in TMDs heterobilayers \cite{bai2020excitons}. In the heterostrained moir\'e superlattices, distinguished IX properties can be expected as compared to the twisting induced ones. A qualitative picture is that the breaking of rotational symmetry can change optical selection rules from circular to linear polarization, but a systematic study of how this happens has not been carried out yet.

In this work we systematically study the mini-band dispersion and optical properties of IXs in heterostrain induced moir\'e, in comparison with those in the moir\'e induced by twisting of various angles. The two types of moir\'e patterns have the valley mismatch between electron and hole in a different manner, which results in distinct distribution of light cones with respect to the exciton kinetic energy dispersion (Fig.~\ref{fig_twistvsstrain}c). The broken rotational symmetry in the strain induced moir\'e manifests as the exchange in positions of the main light cones with different polarized dipoles. Besides, the strain also introduces a constant gauge potential on either electron or hole, which shifts the dispersion of exciton with respect to its crystal momenta in the moir\'e superlattice, leading to dramatic changes of kinetic energies at the main and Umklapp light cones. Upon the band folding by the moir\'e superlattice potential, the mini-band dispersion, wave function spatial profile, and optical properties are totally distinct, even if moir\'e superlattice potentials have the similar real-space profile for the two cases. We also show the evolution of the excitonic mini-bands and the optical dipoles of the bright states inside the light cones with the decrease of moir\'e periodicity, during which the excitonic wave functions evolve from localized wave packets to extended Bloch states. We explore various types of strain configurations and the interplay of twisting and heterostrain, and provide comprehensive diagrams on the optical properties of moir\'e IXs in the strain-parameter space. These studies form the basis of strain and twisting engineering of moir\'e exciton optical properties for potential photonic applications.

The rest of the paper is organized as the following. In Sec.~\ref{section_formalism} we give a brief description of the IX momentum eigenstates, the relation of kinetic energy with respect to its crystal momentum that can be defined from the superlattice Bloch function form, as well as the moir\'e potential that couples the momentum eigenstates. In Sec.~\ref{section_twist}, IXs in twisted heterobilayer are first investigated with the variation of twisting angles to explore the transition between small and large moir\'e regimes. Sec.~\ref{section_strain} presents the strained moir\'e IXs in comparison with the twisting case. The change of BZ geometry and the effective gauge potential from strain are first introduced, and the resultant consequences on the light cone distribution, mini-band dispersion, wave functions, and optical properties are analysed. At last, we briefly discuss properties of IXs under the interplay of twisting and strain. 

\section{IX momentum eigenstate and moir\'e potential\label{section_formalism}}

\subsection{Exciton momentum eigenstates in misaligned bilayers}

TMDs~\cite{liu2015electronic} possess a hexagonal BZ with massive Dirac cones located at the corners $\tau\mathbf{K}$ and $\tau=\pm 1$ is the valley index. Excitons in TMDs exhibit spin-valley locking due to the large spin-orbit splitting. For TMD heterobilayers with type-II band alignment, IXs are composed of electrons and holes located at $\tau'\mathbf{K}'$ and $\tau\mathbf{K}$ valleys in different layers. Here we use prime to denote quantities from the upper layer. Apart from misalignment of the Dirac cones in energy, locations of Dirac cones in momentum space are affected by the configuration of the bilayer. In a twisted bilayer, Dirac cones in the upper layer are rotated with respect to those in the lower layer (Fig.~\ref{fig_twistvsstrain}b upper panel). In a strained bilayer, Dirac cones are first shifted away from the corners of the distorted BZ (red shaded area in Fig.~\ref{fig_twistvsstrain}b lower panel) due to the breaking of three-fold rotational symmetry (empty pink dots). Moreover, the Dirac cones are further translated by a pseudogauge potential caused by changes of the hopping energy (from empty to solid pink dots), which will be discussed in Sec.~\ref{section_formalism:StrainEffects}.

\begin{figure}
	\centering
	\includegraphics[width=0.48\textwidth]{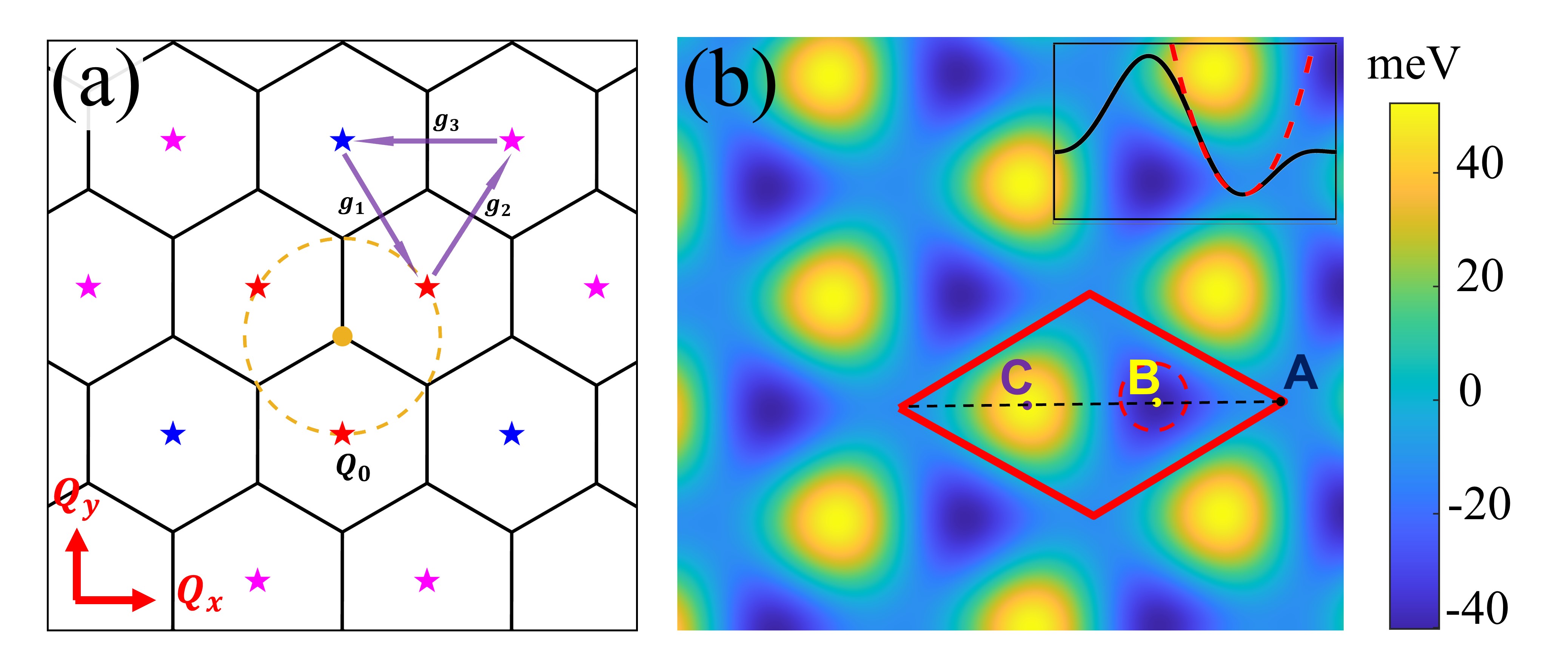}
	
	\caption{(a) Light cones in the extended BZ scheme for a twisted moir\'e. The black hexagons denote the moir\'e BZs, purple arrows mark $\mathbf{g}_1, \mathbf{g}_2$, and $\mathbf{g}_3$. Red stars stand for main light cones, and blue/pink stars represent the 1st/2nd Umklapp light cones. The origin $\mathbf{Q}=0$ is marked by the brown dot and the brown circle is the equi-energy ring from main light cones. 
	(b) The moir\'e potential landscape for R-type $\text{WSe}_2/\text{MoSe}_2$, where red rhombus marks a moir\'e supercell. A, B, and C label three high symmetry locals. The potential inside the red dashed circle can be approximated as a harmonic trap in the small twist angle regime. Upper right inset: Cross-section of the potential along the long axis of the supercell (black solid curve) and the harmonic approximation (red dashed curve).}
	\label{fig_potential}
\end{figure}

Using Bloch wave functions of a pair of electron and hole from each individual layer, $\psi^e_{\mathbf{k}_e}(\mathbf{r}_e)$ and $ \psi^{h}_{\mathbf{k}_h}(\mathbf{r}_h)$, one can construct the IX momentum eigenstate. Since Coulomb interaction conserves the center-of-mass (COM) momentum $\mathbf{Q}$ of electron and hole, it is a good quantum number to characterize the IX momentum eigenstate~\cite{yu2018brightened}:
\begin{eqnarray}
	\label{IXmomentumeigenstate}
& &X_{\tau'\tau,\mathbf{Q}}(\mathbf{R},\mathbf{r}_{eh})\notag\\
&=&  \sum_{\Delta\mathbf{Q}}\Phi(\Delta\mathbf{Q})\psi^e_{\tau'\mathbf{K}'+\frac{m_e}{M_0}\mathbf{Q}+\Delta\mathbf{Q}}\psi^{h*}_{\tau\mathbf{K}-\frac{m_h}{M_0}\mathbf{Q}+\Delta\mathbf{Q}}\notag\\
&=& e^{i(\mathbf{Q}+\tau'\mathbf{K}'-\tau\mathbf{K})\cdot\mathbf{R}}U_{\tau'\tau,\mathbf{Q}}(\mathbf{R},\mathbf{r}_{eh})
\end{eqnarray}
In the above, the coordinates and momenta of electrons and holes have been replaced by their COM and relative motion counterparts $\mathbf{R} = \frac{m_e}{M_0}\mathbf{r}_e+\frac{m_h}{M_0}\mathbf{r}_h$, $\mathbf{r}_{eh} = \mathbf{r}_e-\mathbf{r}_h$, $\mathbf{k}_e = \tau'\mathbf{K}'+\frac{m_e}{M_0}\mathbf{Q}+\Delta\mathbf{Q}$, and $\mathbf{k}_h = \tau\mathbf{K}-\frac{m_h}{M_0}\mathbf{Q}+\Delta\mathbf{Q}$ with $M_0 = m_e+m_h$ the exciton mass, $\Delta\mathbf{Q}$ the relative momentum, and $\Phi(\Delta\mathbf{Q})$ the relative motion wave function.
The COM momentum $\mathbf{Q}$ is also called the \textit{kinetic momentum} since it is associated with the IX's kinetic energy $\hbar^2\mathbf{Q}^2/2M_0$. 
$U_{\tau'\tau,\mathbf{Q}}(\mathbf{R},\mathbf{r}_{eh})$ in the last line of Eq.~(\ref{IXmomentumeigenstate}) is a periodic function built from the periodic parts of the electron and hole's Bloch wave functions~\cite{yu2018brightened}. This makes the IX momentum eigenstate of the Bloch type, where $\mathbf{k}_c = \mathbf{Q}+\tau'\mathbf{K}'-\tau\mathbf{K}$ plays the role of \textit{crystal momentum}. In the following, we will focus on the $\tau=\tau'=+$ valley in R-stacking (or parallel) heterobilayers. The other valley is related by time-reversal symmetry. Also, a consistent coordinate system has been chosen throughout the work, i.e. the zigzag (armchair) crystalline direction as x (y) axis.

Note that the crystal momentum $\mathbf{k}_c$ and kinetic momentum $\mathbf{Q}$ are different for IXs in the moir\'e. This contrasts with intralayer excitons in monolayers, or IXs in aligned lattice-matched heterobilayers, where the two momenta are identical due to $\mathbf{K}'-\mathbf{K}\equiv0$. When an exciton is converted into a photon, momentum conservation requires that the crystal momentum satisfies $\mathbf{k}_c\approx 0$. Therefore, intralayer excitons in monolayers or IXs in aligned lattice-matched heterobilayers have vanishing kinetic momentum (brown dot in Fig.~\ref{fig_potential}a). In contrast, bright IXs in the moir\'e have finite kinetic momentum $\mathbf{Q}_{lc}=\mathbf{K}-\mathbf{K}'+\mathbf{G}$   (stars in Fig.~\ref{fig_potential}a), where $\mathbf{G}$ denotes the moir\'e reciprocal lattice vector. $\mathbf{Q}_{lc}$ defines the location of moir\'e light cones, inside which the direct conversion with photon is permitted (Fig.~\ref{fig_twistvsstrain}b).

Moir\'e IX momentum eigenstates in various light cones possess distinct optical dipoles coupled to different elliptically polarized light~\cite{yu2015anomalous} (Fig.~\ref{fig_twistvsstrain}c). In the case of twisted heterobilayer moir\'e, the three equivalent innermost light cones are located at $\mathbf{Q}_0=\mathbf{K}-\mathbf{K}'$, $C_3\mathbf{Q}_0$, and $C^2_3\mathbf{Q}_0$ (red stars in Fig.~\ref{fig_potential}a, $C_3$ denotes three-fold rotation). These are the main light cones, in which the momentum eigenstates hold dominant optical dipoles. Farther away are the Umklapp light cones, which represent the Umklapp recombination process with much weaker dipoles. Such categorization can also be applied to strained heterobilayer moir\'e, although three-fold rotational symmetry is broken.

\subsection{Moir\'e potential}

Moir\'e patterns from the misaligned bilayers exhibit spatially modulated local atomic registry that repeats in a much larger scale than the monolayer lattice constant. 
Different local atomic configurations in a moir\'e render local-to-local variation on the interlayer vdW interaction, forming a lateral modulation on the local band gap and interlayer distance~\cite{yu2018brightened}. The IX thus experiences a moir\'e potential, which can be modeled by~\cite{wang2017interlayer,yu2017moire,wu2018theory,yu2020electrically} 
\begin{eqnarray}
\label{moirepotentialformula}
V(\mathbf{R}) = \sum_{n = 1}^3 2V_0\cos\left(\mathbf{g}_n\cdot\mathbf{R}-\varphi\right),
\end{eqnarray}
where $\mathbf{g}_1$, $\mathbf{g}_2$, and $\mathbf{g}_3 = -\mathbf{g}_1-\mathbf{g}_2$ are the primitive reciprocal lattice vectors of the moir\'e superlattice separated by $120^\circ$ (purple arrows in Fig.~\ref{fig_potential}a). The information of the moir\'e profile is contained in the reciprocal lattice vectors, which vary for different moir\'e superlattices (e.g. the two cases in Fig.~\ref{fig_twistvsstrain}). Values of $V_0$ and $\varphi$ depend on materials and their stacking configuration~\cite{wu2017topological,wu2018theory}. For example, $V_0 = 9.122$ meV and $\varphi \approx 0.57\pi$ for R-type $\text{WSe}_2/\text{MoSe}_2$ heterobilayer~\cite{yu2017moire,yu2020electrically}. The potential profile in a twisted moir\'e superlattice is shown in Fig.~\ref{fig_potential}b. Each potential minimum (B) is connected with three saddle points (A) and three maxima (C). The three locals A, B, and C correspond to $R_h^h$, $R_h^X$, and $R_h^M$ high symmetry stacking registries, respectively (Fig.~\ref{fig_twistvsstrain}a). Here, $R_h^\mu$ denotes R-type stacking, with the $\mu$ site of the electron layer vertically aligned with the hexagon center $(h)$ of the hole layer. $M$ and $X$ represent metal and chalcogen atoms. 

The momentum space Hamiltonian describing the moir\'e exciton Bloch states in the slowly varying moir\'e potential reads~\cite{yu2017moire,yu2020electrically}
\begin{eqnarray}
	\label{twistHamiltonian}
	H &=& \sum_l \left[\left(E_X+\frac{\hbar^2 \vert \mathbf{Q}_l \vert ^2}{2M_0}\right) \ket{\mathbf{Q}_l}  \bra {\mathbf{Q}_l} \right]\nonumber\\
	&+& \sum_l \left[\sum_{n=1}^3(V_0 \ket{\mathbf{Q}_l+\mathbf{g}_n}  \bra {\mathbf{Q}_l} +h.c.)\right]
\end{eqnarray}
where $\ket{\mathbf{Q}_l}$ is the momentum eigenstate in the $l$th mini BZ, $E_X\approx 1.40$ eV and is tunable with electric field. The second line illustrates that each momentum eigenstate is coupled to six neighboring states with momentum difference $\pm\mathbf{g}_n$, which will lead to the formation of mini-bands.

\begin{figure}
	\centering
	\includegraphics[width=0.48\textwidth]{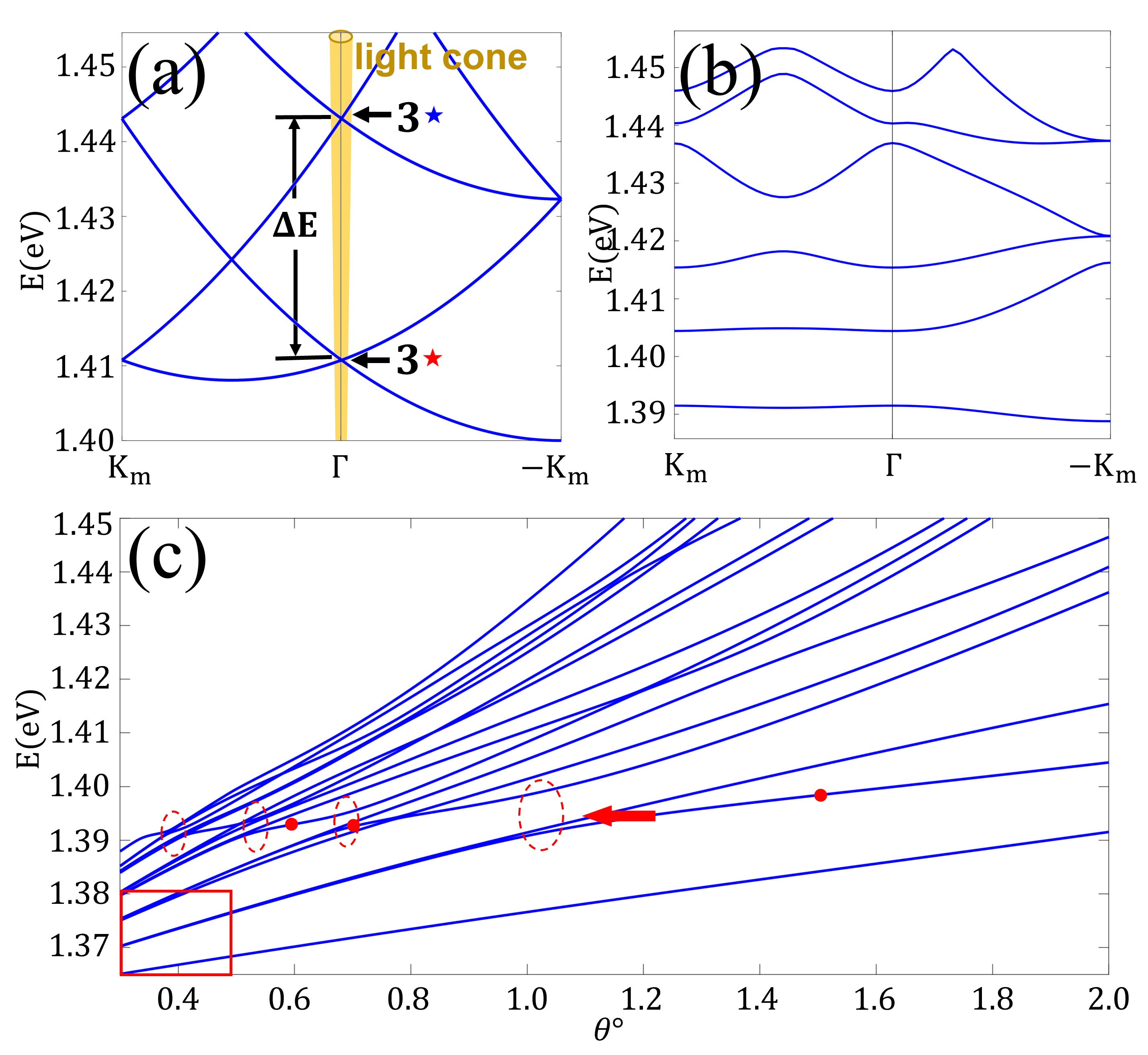}

	\caption{Dispersion relations of IX in twisted heterobilayer moir\'e. (a) The dispersion in the mini BZ at $\theta = 2^\circ$ in the absence of moir\'e potential. $\pm\mathbf{K}_m$ are the mini BZ corners. There are two degenerate points at $\mathbf{\Gamma}$, which consist of three main light cones and three 1st Umklapp light cones, respectively. The yellow cone stands for the light cone with very sharp slope, so only the states near $\mathbf{\Gamma}$ point can couple to light directly. (b) The dispersion in the presence of moir\'e potential. Gaps are opened at the degenerate points in (a). (c) The first 16 energy levels at $\mathbf{\Gamma}$ with the variation of twist angle from $0.3^\circ$ to $2.0^\circ$. States inside the red box exhibit equal energy spacing and linear scaling. There is a plateau state around $1.39$ eV, whose energy barely changes with $\theta$. Red ellipses mark the hybridization of this state with other states forming anti-crossings. The red dots mark the states whose wave functions are shown in Fig.~\ref{fig_twistexciton}b.}
	\label{fig_twistband}    
\end{figure}

\begin{figure*}
	\centering
	\setlength{\abovecaptionskip}{0.cm}
	\includegraphics[width=1\textwidth]{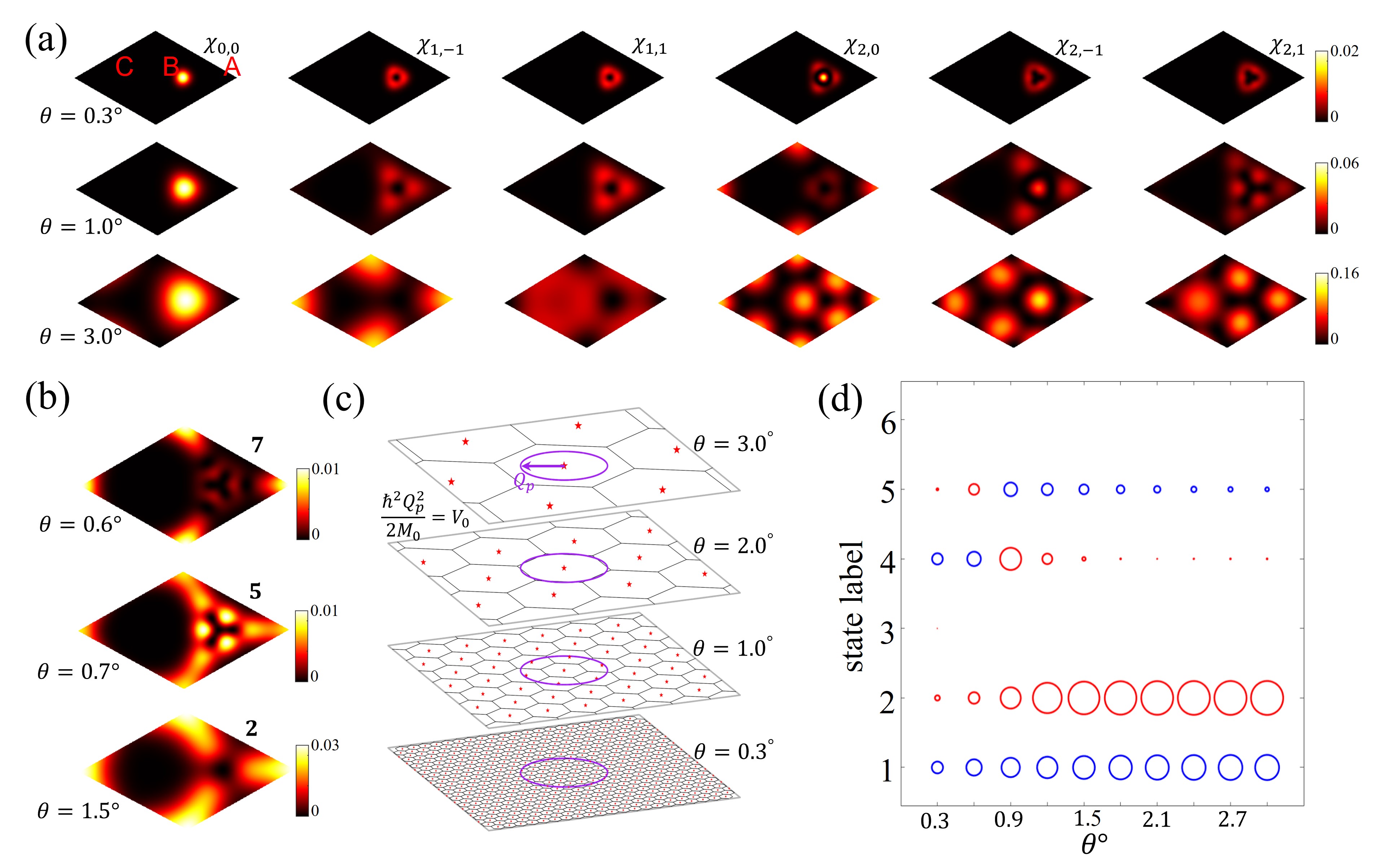}
	
	\caption{Wave function distribution for the low energy exciton mini-bands in twisted heterobilayer moir\'e of various twist angles. (a) Wave function densities of the first 6 states at $\theta = 0.3^\circ, 1.0^\circ$, and $3.0^\circ$. Wave functions at small twist angle possesses harmonic forms, where the orbital symmetry is labeled on the first row. (b) Evolution of the plateau state with twist angle (c.f. Fig.~\ref{fig_twistband}c, where the corresponding states are marked by red dots). It mainly distributes around A locals and will hybridize with states located around B point if their energies intersect ($\theta = 0.7^\circ$). (c) The effective coupling range dictated by moir\'e potential strength. Light cone (red stars) distribution in the extended mini-BZs is shown at different twist angles, where the circle has radius $Q_p \equiv \frac{\sqrt{2mV_0}}{\hbar}$, subscript p stands for moir\'e potential. Momentum eigenstates within the range characterized by $Q_p$ are strongly coupled by the moir\'e potential. (d) In-plane polarization of moir\'e excitons for $\theta = 0.3^\circ$ to $3.0^\circ$. Red/blue color stands for right/left circular polarization, circle size stands for dipole amplitude.}
	\label{fig_twistexciton}
\end{figure*}

\section{Interlayer excitons in  twisted heterobilayer moir\'e\label{section_twist}}

We first investigate IXs in twisted heterobilayer moir\'e with various twisting angles. We take R-type $\text{WSe}_2/\text{MoSe}_2$ bilayer and consider the spin singlet moir\'e exciton~\cite{yu2018brightened}.

\subsection{Energy dispersion} 

Figs.~\ref{fig_twistband}a,b show the IX dispersion along $-\mathbf{K}_m$-$\mathbf{\Gamma}$-$\mathbf{K}_m$ in the mini BZ without and with the moir\'e potential at twist angle $\theta=2^\circ$. In the absence of moir\'e potential, the dispersion only comes from the parabolic kinetic energy. $\mathbf{Q}_{\text{lc}}$, where light cones reside, are folded onto the $\mathbf{\Gamma}$ point. The first energy level at $\mathbf{\Gamma}$ is three-fold degenerate consisting of three main light cones (red stars in Fig.~\ref{fig_potential}a). The second level is composed of three 1st Umklapp light cones (blue stars in Fig.~\ref{fig_potential}a). In the presence of moir\'e potential, degeneracy is broken with gap opening. If the energy separation between the two levels is smaller than the potential energy, i.e., $\Delta E < V_0$ (Fig.~\ref{fig_twistband}a), coupling between main and 1st Umklapp light cones will occur~\cite{yu2020electrically}.

With the variation of twist angle, energy levels at the $\mathbf{\Gamma}$ point will evolve and certain states may mix (Fig.~\ref{fig_twistband}c). The kinetic momentum $\mathbf{Q}$ increases approximately linearly with $\theta$, thus the kinetic energy evolves with $|\mathbf{Q}|^2 \sim \theta^2$. In the small angle regime ($\theta < 1.5^\circ$), the separation of kinetic energy between the main and 1st Umklapp light cones is smaller than or comparable to the moir\'e potential strength, i.e., $\Delta E\lesssim V_0$. Thus, strong coupling between main and Umklapp light cones is expected. When twist angle is large ($\theta \gtrsim 3.0^\circ$), $\Delta E > V_0$, the main and Umklapp light cones are effectively decoupled.

Several interesting features can be identified in the evolution of energy levels with twist angle. First, the lowest a few levels exhibit equal spacing and scale linearly when $\theta$ is small (red box in Fig.~\ref{fig_twistband}c). Second, there is a plateau state around $1.39$ eV whose energy is robust against variation of $\theta$. It corresponds to the 16th state at $\theta=0.35^\circ$ and becomes the 2nd when $\theta>1.1^\circ$ since the other levels increase with $\theta$ and rise to higher energies.

\subsection{Wave function distribution}

It is also interesting to look at the wave function distribution for the first a few states at $\mathbf{\Gamma}$ point (Fig.~\ref{fig_twistexciton}a). One can notice that the first a few states are localized around potential minima (B point) at small angles, e.g., $\theta = 0.3^\circ$. Particularly, profiles of the first a few states at small angles are analogous to the eigenstates of a two-dimensional harmonic oscillator with the constraint of three-fold rotational symmetry. Consequently, we can label the states as $\chi_{n,l}$, where $n$ is the principal quantum number, and $l$ is the angular momentum quantum number modulo 3 (1st row of Fig.~\ref{fig_twistexciton}a). These features and the scaling of energy levels discussed in the previous paragraph can be easily understood by realizing that the low energy IXs are trapped at the potential minima (B point), which can be well approximated as a harmonic oscillator potential (Fig.~\ref{fig_potential}b inset) $V(\rho)\approx 1.47V_0\left(4\pi^2\rho^2/a_m^2-3\right)$ with $\rho$ the radial distance measured from B point and $a_m$ the moir\'e period. Since the frequency of the harmonic oscillator $\omega \propto a_m^{-1}\propto \theta$, one recovers the equal spacing and linear in $\theta$ scaling at low energies. When $\theta$ is enlarged, high symmetry locals in the moir\'e shrink rapidly and the hopping between adjacent trapping sites in different supercells emerge, thus wave functions start to spread (2nd row of Fig.~\ref{fig_twistexciton}a). For even larger twist angles, the harmonic oscillator approximation breaks down, for example, the first three states mainly occupy B, A, C locals respectively when $\theta = 3.0^\circ$ (3rd row of Fig.~\ref{fig_twistexciton}a). Such wave function evolution can also be understood from a momentum space perspective. Momentum eigenstates in neighboring moir\'e BZs are coupled by moir\'e potential, whose strength can be characterized by an effective momentum length $Q_p$ satisfying $\hbar^2Q_p^2/2M_0 = V_0$. The purple circles in Fig.~\ref{fig_twistexciton}c delimit the region defined by the effective momentum. At small twist angles, the circle covers a large number of light cones. Many momentum eigenstates participate in constructing a moir\'e exciton eigenstate, the strong coupling among which yields a localized wave function. At large twist angles, the separation between light cones is increased, rendering a dramatically reduced number of light cones inside this circle. The wave functions of the excitons are built with less coupled momentum eigenstates and inherit their extended Bloch wave nature.

Now let us look at some features of the plateau state around $1.39$ eV. It corresponds to the second state when $\theta > 1.1^\circ$, while its ordering changes with smaller $\theta$ and hybridization with other states occur around certain angles. It mainly distributes around the A high symmetry locals at most twist angles as shown by the top and bottom panels of Fig.~\ref{fig_twistexciton}b. This helps understand its insensitivity to variation of $\theta$: Excitons residing around A locals experience a rather flat potential landscape with magnitude around $-V_0$, whose energy barely changes with the size of moir\'e. Therefore, the plateau state exhibits an almost constant energy around $E_X-V_0\approx1.39$ eV. At twist angles where the energy level of the plateau state intersects with others, strong hybridization occur between it and those states located around B with the appearance of avoided crossings as marked by the dashed circles in Fig.~\ref{fig_twistband}c. For example, at $\theta=0.7^\circ$ (middle panel of Fig.~\ref{fig_twistexciton}b), the plateau state hybridizes with an extended $W_{2,-1}$ state rendering large densities around the B local.

\subsection{Optical properties}

Next we investigate the optical properties of twist moir\'e IXs at various angles. We focus on the first six states at $\mathbf{\Gamma}$ (Fig.~\ref{fig_twistband}b). One finds that all the states couple with circularly polarized light except the 3rd and 6th states, whose in-plane dipole vanishes (Fig.~\ref{fig_twistexciton}d). Red/blue color represents right/left circular direction, the size of circles denote the dipole amplitude. When $\theta < 1.5^\circ$, $\Delta E\lesssim V_0$ and the effective momentum $Q_p$ that characterizes the range of moir\'e potential covers many light cones (Fig.~\ref{fig_twistexciton}c), states in the main and 1st Umklapp light cones couple significantly. Thus, strong optical dipoles from main light cones and weak dipoles from Umklapp light cones are mixed, producing comparable dipole strengths for different states. The insensitivity of the plateau state energy against variation of $\theta$ is also reflected on optical properties. The switching of polarization in the 4th and 5th states from $\theta = 0.6^\circ$ to $0.9^\circ$ is related to the reordering of plateau state-- It corresponds to the 5th state at $\theta = 0.6^\circ$, then exchanges order with its neighbor and becomes the 4th state at $\theta = 0.9^\circ$. 
With the twist angle enlarged, the coupling between main and 1st Umklapp light cones becomes weaker. Optical dipoles of the first three states are mostly contributed by the three main light cones. The fourth to sixth states are mainly governed by the three 1st Umklapp light cones with much weaker dipole intensities. At $\theta = 3.0^\circ$, wave functions for the three lowest states with dominating dipoles are well separated (3rd row of Fig.~\ref{fig_twistexciton}a):
The 1st to 3rd states are centered at B, A, and C locals, respectively. Consequently, A, B locals in the moir\'e supercells are coupled uniquely to the $\sigma_+$, $\sigma_-$ polarized light, respectively, and in-plane polarization is forbidden at C locals (Fig.~\ref{fig_twistvsstrain}a upper panel)~\cite{yu2017moire, yu2018brightened}. The 4th to 6th states have much wider spread, however, the location-dependent optical properties remain applicable. For instance, the 4th (5th) state contributes $\sigma_+$ ($\sigma_-$) light emitting at the A (B) local (among other places), although the intensity is much weaker.

The observed optical properties in both small and large angle regime can be understood through symmetry analysis. The photon polarization from optical recombination of moir\'e excitons depends on the rotational symmetry. At small twist angles, the photon polarization is dictated by symmetry of local orbitals of the wave functions (first row of Fig.~\ref{fig_twistexciton}a). The moir\'e exciton in a wide potential well can be approximated as a wave packet \cite{yu2017moire} $\chi_{n,l} = \sum_{\mathbf{Q}} e^{-i(\mathbf{Q}-\mathbf{Q}_0) \cdot \mathbf{R}_c} W_{n,l}(\mathbf{Q}) X_{\mathbf{Q}}$, where $W_{n,l}(\mathbf{Q})$ is the 2D harmonic envelope wave function constraint by three-fold rotational symmetry. Setting the wave packet center $\mathbf{R}_c$ at the B point, the in-plane optical dipole $\mathbf{D}_\parallel$ can be analyzed via
\begin{eqnarray}
	\label{localorbitaldipole}
	\hat{e}_+ \cdot \mathbf{D}_\parallel &\sim& W_{n,l}(\mathbf{Q}_0)+e^{i\frac{2\pi}{3}} W_{n,l}(C_3 \mathbf{Q}_0)+e^{i\frac{4\pi}{3}}W_{n,l}(C_3^2 \mathbf{Q}_0)\notag\\
	\hat{e}_- \cdot \mathbf{D}_\parallel &\sim& W_{n,l}(\mathbf{Q}_0)+ W_{n,l}(C_3 \mathbf{Q}_0)+W_{n,l}(C_3^2 \mathbf{Q}_0)
\end{eqnarray}
For example, at $\theta = 0.3^\circ$, $W_{0,0}(\mathbf{Q})=W_{0,0}(Q)$ is a real function of $Q$ (Fig.~\ref{fig_twistexciton}a). This yields $\hat{e}_+ \cdot \mathbf{D}_\parallel \sim W_{0,0}(Q_0)+e^{i\frac{2\pi}{3}} W_{0,0}(Q_0)+e^{i\frac{4\pi}{3}} W_{0,0}(Q_0)=0$, while $\hat{e}_- \cdot \mathbf{D}_\parallel \sim W_{0,0}(Q_0)+ W_{0,0}(Q_0)+W_{0,0}(Q_0)$ is finite. By the same token, the above orbital optical selection rule determines that states with $l = 0$ emit $\sigma_-$ light, states with $l = -1$ emit $\sigma_+$ light, while states with $l = 1$ have vanishing in-plane polarization, which is consistent with results of Fig.~\ref{fig_twistexciton}d (e.g., first column).

At large angles (e.g., $\theta = 3.0^\circ$), momentum eigenstates in neighboring BZs are weakly coupled, so wave functions become more extended. The optical dipoles instead depend on real-space atomic configurations. The electron and hole Bloch functions have distinct $C_3$ eigenvalues about different rotation centers ($h$ center, chalcogen site, or metal site, Fig.~\ref{fig_twistvsstrain}a middle panel). Moir\'e excitons $\chi$ at different high symmetry locals have distinct rotation centers, thus they exhibit location-dependent $C_3$ transformations~\cite{yu2018brightened,yu2017moire}
\begin{eqnarray}
	\label{C3transofmration}
	C_3\chi_A = e^{-i\frac{2\pi}{3}}\chi_A,
	C_3\chi_B = e^{i\frac{2\pi}{3}}\chi_B,
	C_3\chi_C = \chi_C 
\end{eqnarray}
Photons converted from such excitons possess the same symmetry. Since the first six states have specific rotational centers, moir\'e excitons in the large angle regime exhibit location-dependent optical selection rules as schematically shown in Fig.~\ref{fig_twistvsstrain}a.

\section{Interlayer excitons in heterostrained moir\'e\label{section_strain}}

In this section we consider IXs in moir\'e formed by various types of heterostrain~\footnote{Strain can be induced by various methods, such as pulling on suspended sheets with an electrostatic gate, or by bending a flexible substrate\cite{frisenda2017biaxial,roldan2015strain,deng2018strain,yang2021strain,han2021experimental}. In experiments, errors/uncertainties may occur when applying strain to the samples. First, not all the applied strain to the substrate was transferred to the sample~\cite{liu2014strain,li2020efficient}. This uncertainty can be calibrated by measuring strain transfer efficiency. Besides, strain distribution might be nonuniform due to interactions between the substrate and the sample. Experimentally, strain with uncertainty $<0.1\%$ in a few microns’ scale can be achieved~\cite{liu2014strain,goldsche2018tailoring}. The optical properties of interlayer excitons usually have no qualitative changes within strain variation of $< 0.1\%$ (Fig.~\ref{fig_generalstrain}, with exceptions near transition boundaries).} For simplicity, we assume that only the top layer is stained. We compare properties of IXs in strained moir\'e with those in the twisted case.
 
\subsection{Strain effect}\label{section_formalism:StrainEffects}

To characterize the IX in a moir\'e formed by heterostrain, first we give a brief introduction to the geometric and electronic effects of strain~\cite{vozmediano2010gauge,gerardo2017electronic,fang2018electronic,rostami2015theory}. 

First, strain introduces geometric changes to the lattice structure (Fig.~\ref{fig_twistvsstrain}a lower panel). Such geometric variations shift the location of Dirac cones to  $\tau(I+S)^{-1}\mathbf{K}$ (dashed pink circles in Fig.~\ref{fig_twistvsstrain}b lower panel), where $I$ is the identity matrix and $S$
is the strain tensor.

Apart from geometric effects on the crystalline structure, strain also modifies the hopping energy along different directions. In the limit of small strain, such variation can be captured by a pseudogauge potential in terms of strain tensor components
\begin{eqnarray}
\mathbf{A} = \frac{\sqrt{3}}{2a}\beta(\epsilon_{xx}-\epsilon_{yy},-2\epsilon_{xy})^T,
\end{eqnarray}
where $\beta$ is a material-dependent parameter, for instance, $\beta \approx 2.30$ for WSe$_2$~\cite{fang2018electronic}. Consequently, the two Dirac points in the monolayer BZ are further shifted oppositely (to respect time-reversal symmetry) by the pseudogauge potential towards 
\begin{eqnarray}
\tau\mathbf{D} = \tau(I+S)^{-1}\mathbf{K} -\tau \mathbf{A},
\end{eqnarray}
as shown schematically from pink circles to pink dots in Fig.~\ref{fig_twistvsstrain}b lower panel. The first term represents the distorted Dirac points due to geometric distortion discussed in the previous paragraph. In contrast to the twisting case, the Dirac points are shifted away from the BZ corners.

\subsection{Volume-preserving heterostrain}

\begin{figure}
	\centering
	\includegraphics[width=0.48\textwidth]{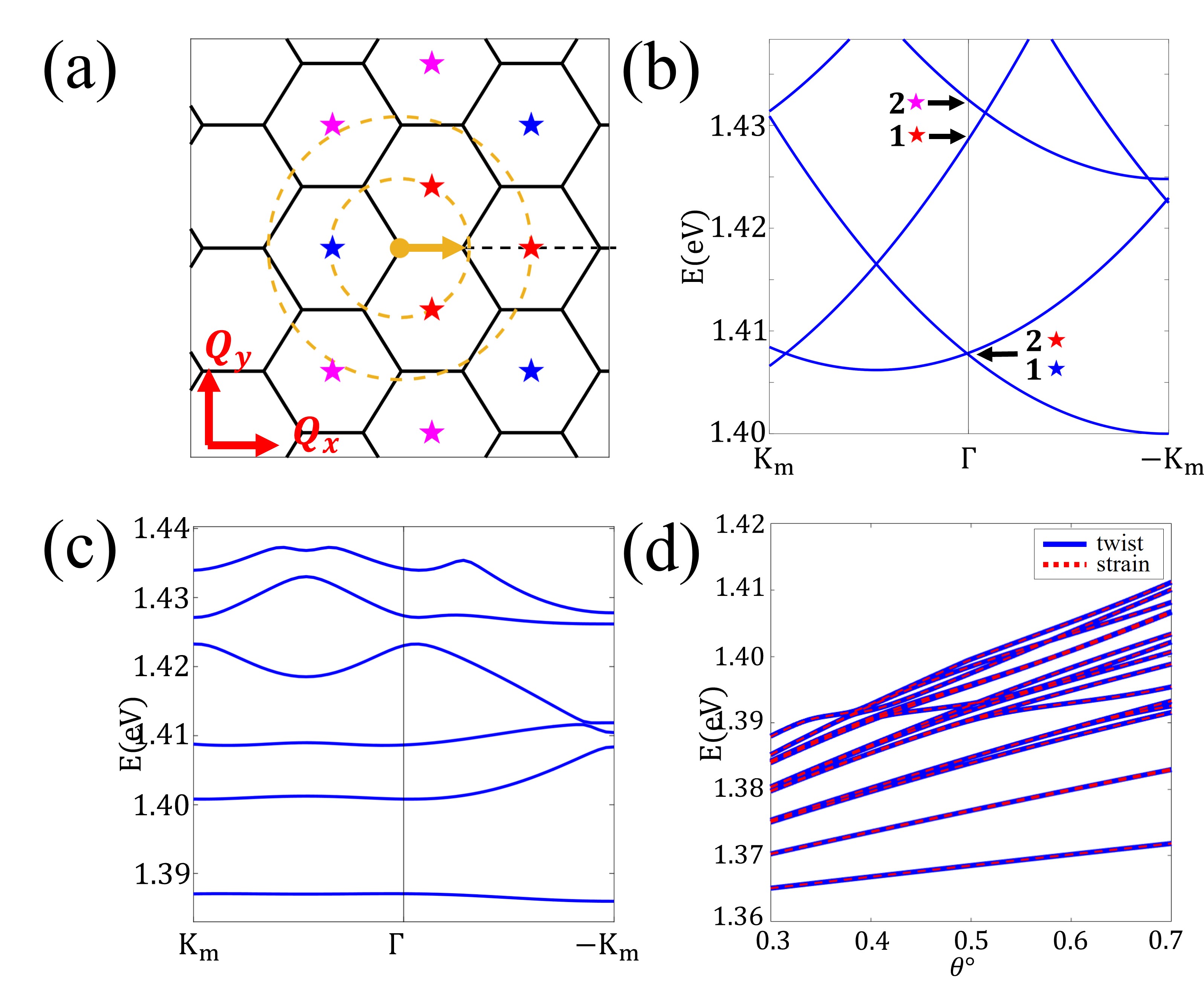}
	
	\caption{(a) Light cone distribution in the moir\'e formed by volume-preserving strain along the zigzag direction. Black hexagons denote the moir\'e BZs. Red and blue/pink stars represent main and 1st/2nd Umklapp light cones. The brown dot, arrow and dashed circles represent the origin, gauge potential $\mathbf{A}$, and equi-energy rings, respectively. (b) The kinetic energy dispersion of heterostrained moir\'e excitons at $\epsilon = 3.0\%$. Three-fold degeneracies at $\mathbf{\Gamma}$ are broken due to strain. Composition of light cones for different levels at $\mathbf{\Gamma}$ are also labeled. (c) Dispersion of heterostrained moir\'e excitons in the presence of moir\'e potential. (d) Comparison of energy levels at $\mathbf{\Gamma}$ between twisted (blue) and heterostrained (red dashed) moir\'e excitons. Note that strain intensity has been converted into angles to facilitate the comparison (i.e. $\epsilon/\pi \times 180^\circ$).}
	\label{fig_strainband}    
\end{figure}

\begin{figure*}[t]
	\centering
	\includegraphics[width=1\textwidth]{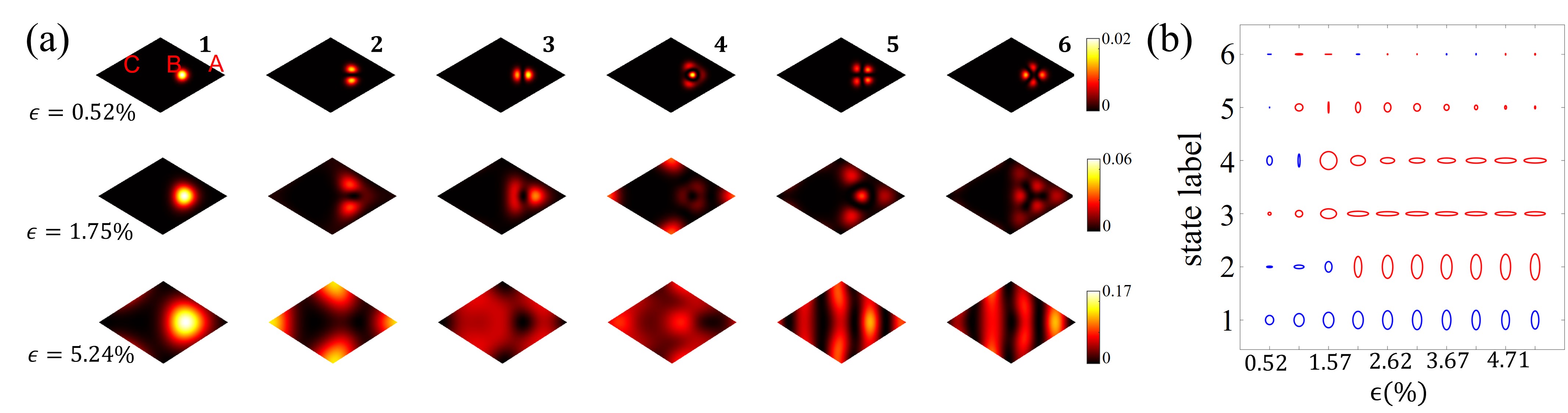}
	
	\caption{(a) Wave functions of the first 6 lowest states of moir\'e excitons in volume-preserving heterostrained moir\'e along the zigzag direction. The wave functions' three-fold rotational symmetry is broken, while the mirror symmetry along the long-axis of the moir\'e unit cell remains. (b) The in-plane polarization of strain moir\'e excitons from $\epsilon = 0.52\%$ to $5.2\%$. Red/blue color stands for positive/negative helicities, ellipse size stands for dipole amplitude.}
	\label{fig_strainwavefunc}
\end{figure*}

Here we consider volume-preserving strain~\cite{bi2019designing,shabani2021deep}, i.e., stretching the top layer along the zigzag direction while compressing it with the same extent in the perpendicular direction (Fig.~\ref{fig_twistvsstrain}a). The strain tensor reads $S = \text{diag}(\epsilon,-\epsilon)$, where $\epsilon$ is the strain strength. In this case, the large-scale moir\'e landscape resembles that of a twisted bilayer (Fig.~\ref{fig_twistvsstrain}a). This similarity also applies to the superlattice potentials based on the local approximation for long-period moir\'e (Appendix~\ref{section_similarity}). This allows us to focus first on the effects of the pseudo-gauge potential by studying the volume-preserving strain. From discussions in Sec.~\ref{section_formalism:StrainEffects}, we know that strain shifts Dirac cones away from BZ corners. Thus, the kinetic momentum in Eq.~(\ref{twistHamiltonian}) should be replaced by
\begin{eqnarray}
\mathbf{Q}_l^s = \mathbf{Q}_l+\mathbf{A}
\end{eqnarray}
in the presence of strain. Fig.~\ref{fig_strainband}a shows the distribution of light cones in volume-preserving strained moir\'e, where the brown arrow denotes the shift caused by the gauge potential $\mathbf{A}$ with respect to the origin (brown dot). Interestingly, here the size of $\mathbf{A}$ is almost identical to the length of the mini BZ boundary, which results in the following redistribution of the light cones. $C_3$ symmetry of the three main light cones is broken, which is manifested on the exchange of the dipole polarization of the two main light cones on the two sides of the yellow arrow in Fig.~\ref{fig_twistvsstrain}c~\cite{yu2015anomalous}. In contrast to the twisting case in Fig.~\ref{fig_potential}a, the three main light cones are no longer sitting on the same equi-energy circle (smaller ring in Fig.~\ref{fig_strainband}a), with one of them shifted to higher energy (larger ring). Meanwhile, one of the 1st Umklapp light cones moves towards the origin and stays close to the smaller equi-energy circle. As will be shown later, this change will suppress the optical strength of the low energy states.

Now we look at the effects of heterostrain on the dispersion of IXs. Fig.~\ref{fig_strainband}b shows the exciton dispersion without the moir\'e potential along the direction marked by the black dashed line in Fig.~\ref{fig_strainband}a. The first (approximate) degeneracy at $\mathbf{\Gamma}$ consists of two degenerate main light cones and a 1st Umklapp light cone with slightly different energy. The second level at $\mathbf{\Gamma}$ originates from the remaining main light cone and the third level comprises of two degenerate 2nd Umklapp light cones (pink stars near the larger ring in Fig.~\ref{fig_strainband}a). Such redistribution of main and Umklapp light cones among the states, as compared to the case of twisting (Fig.~\ref{fig_twistband}a), will yield distinct optical properties.

\begin{figure}
	\centering
	\includegraphics[width=0.48\textwidth]{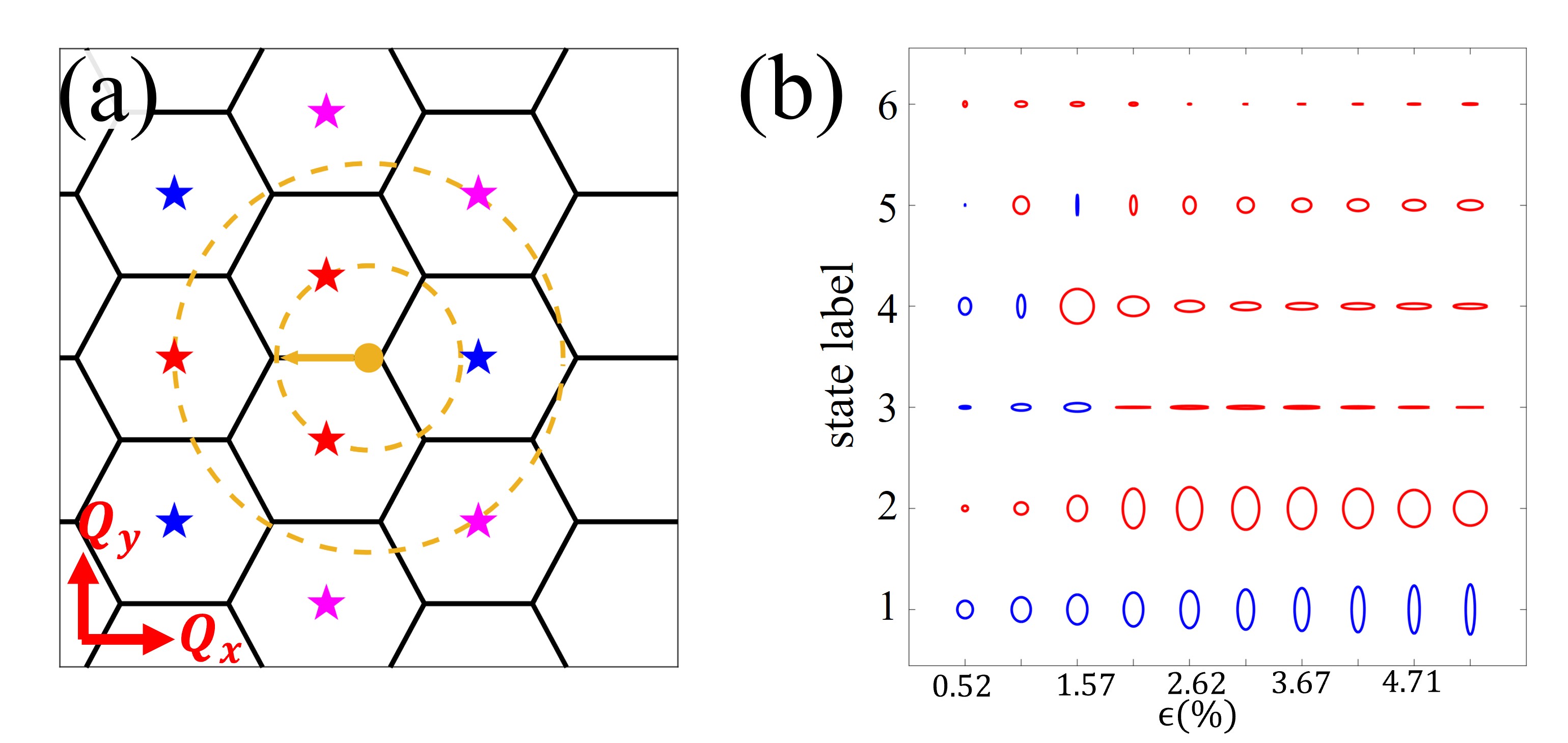}
	
	\caption{Moir\'e excitons from volume-preserving heterostrain along the armchair direction. (a) The light cone distribution. Black hexagons denote the moir\'e BZs. Red and blue/pink stars represent main and 1st/2nd Umklapp light cones. The brown dot, arrow and dashed circles represent the origin, gauge potential $\mathbf{A}$, and equi-energy rings, respectively. (b) The in-plane polarization of strain moir\'e excitons from $\epsilon = 0.52\%$ to $5.2\%$. Red/blue color stands for positive/negative helicities, ellipse size stands for dipole amplitude.}
	\label{fig_strain90}
\end{figure}

For different types of moir\'e superlattices, e.g., twisting vs heterostrain, the moir\'e potential is only affected via changes of moir\'e primitive reciprocal lattice vectors (Eq.~(\ref{moirepotentialformula})). This only changes the spatial profile instead of the magnitude of the moir\'e potential. Analogous to the twisting case, momentum eigenstates in different moir\'e BZs are coupled through the Fourier components of the moir\'e potential. On the other hand, the kinetic energy is affected by the strain-induced pseudogauge potential and -- different from the twisting case -- $\mathbf{Q}_l$ should be replaced by $\mathbf{Q}_l^s$ in Eq.~(\ref{twistHamiltonian}) for strained moir\'e excitons. The mini-bands in the presence of moir\'e potential is shown in Fig.~\ref{fig_strainband}c, which resembles that of the twisted moir\'e (Fig.~\ref{fig_twistband}b).

Such similarity extends to the evolution of energy levels at $\mathbf{\Gamma}$ versus the variation of $\epsilon$ or $\theta$ (Fig.~\ref{fig_strainband}d). This is because energy levels only depend on strength of moir\'e potential and the distribution of light cones irrespective of their nature (main or Umklapp)\footnote{The similarity of $A$ and $K_m$ is caused by the material-dependent parameter $\beta$. For volume-preserving strain, the $A = \frac{\sqrt{3}\beta}{2a}(\epsilon_{xx}-\epsilon_{yy})=\frac{\sqrt{3}\beta \epsilon}{a}$, and $K_m = \frac{4\pi}{3a_m}\approx \frac{4\pi\epsilon}{3a}$, where $a_m\approx a/\epsilon$ is moir\'e period. The ratio of the two quantities becomes $\frac{A}{K_m} = \frac{3\sqrt{3}\beta}{4\pi}$, which only depends on $\beta$. The ratio is approximately 0.95 (0.99) for $\text{WSe}_2$ ($\text{MoSe}_2$) whose $\beta=0.23\,(0.24)$.}.

Instead, different distributions of light cones between the heterostrain and twisting cases are reflected on the wave functions (Fig.~\ref{fig_strainwavefunc}a). In general, the local densities of the wave functions lose three-fold rotational symmetry. However, the mirror symmetry with respect to the long axis of the supercell is retained. Despite changes in symmetry, the features of spatial distribution remain unchanged: For small strain strength, wave functions exhibit local orbital features at potential minima; For large strain strength, wave functions exhibit extended Bloch wave forms, where quasi-1D stripe patterns can be spotted for high energy states.

The optical properties of strained moir\'e excitons are also remarkably distinct from the twisted ones (Fig.~\ref{fig_strainwavefunc}b). Compared with twisted excitons, one finds: (i) The first four states at $\mathbf{\Gamma}$ possess comparable dipole strength independent of strain intensity. While for twisted moir\'e excitons, the first two states possess prominent dipole strength, especially at large twist angles (Fig.~\ref{fig_twistexciton}d); (ii) The 3rd and 6th states have finite dipole strength in contrast to the vanishing contributions in the case of twisting; (iii) The coupled light in strained moir\'e is elliptically instead of circularly polarized. The first difference results from the light cone redistribution. As is shown in Fig.~\ref{fig_strainband}b, the lower states at $\mathbf{\Gamma}$ point are composed of two main light cones and one 1st Umklapp light cone, while the higher states comprise of two 2nd Umklapp light cones and one main light cone. The average effects of strong and weak dipoles from main and Umklapp light cones make the first 4 states exhibit comparable dipole strengths. The second and third differences can be attributed to the distinct symmetry properties. The breaking of three-fold rotational symmetry leads to optical dipole with left and right circular polarization of different strengths, thus forming elliptically polarized light. Notice that the 3rd state still locates around C locals at $\epsilon = 5.24\%$ (Fig.~\ref{fig_strainwavefunc}a), however, the symmetry breaking at C locals ensures that in-plane polarized light-matter coupling is permitted, in contrast to the twisting case. While for the 6th state, its strip-like distribution spread across different locals, which leads to the finite in-plane dipole as compared to its vanishing counterpart in twisted moir\'e. Similar to the case of twisting, the change of the ordering of the plateau state is also reflected in the optical properties between the 4th and 5th states for strained moir\'e excitons.

One can also apply the volume-preserving strain along the armchair direction with the strain tensor $S = \text{diag}(-\epsilon,\epsilon)$. The light cone distribution exhibits an approximate mirror reflection in the $\mathbf{Q}_y$ direction compared with the above case (Fig.~\ref{fig_strain90}a vs Fig.~\ref{fig_strainband}a). Such relation in the light cone distribution guarantees that the optical dipoles in the two cases are similar (Fig.~\ref{fig_strain90}b vs Fig.~\ref{fig_strainwavefunc}b).

\subsection{Uniaxial strain}

\begin{figure*}[t]
	\centering
	\includegraphics[width=1\textwidth]{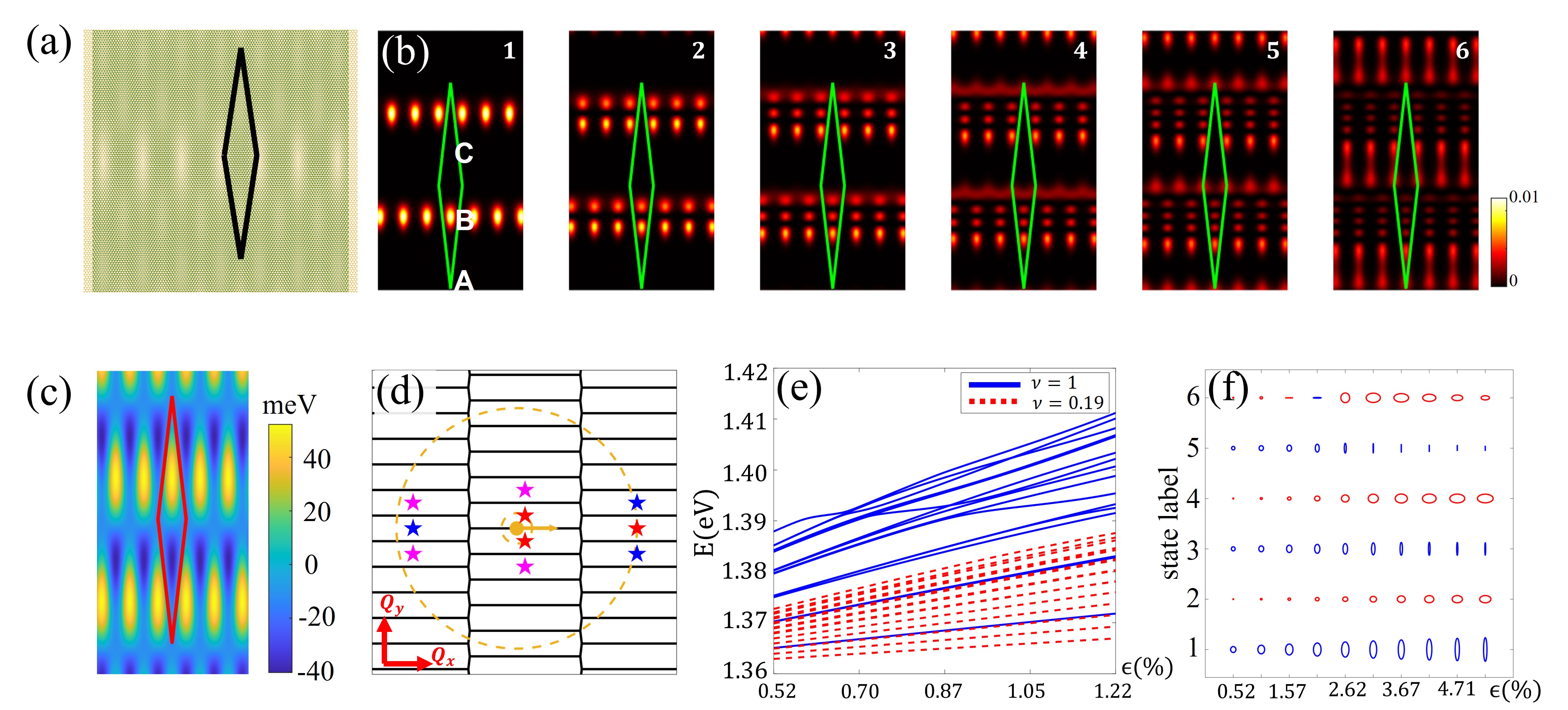}
	
	\caption{Moir\'e excitons from uniaxial heterostrain. (a) Real-space lattice of a heterobilayer, where yellow/green atoms are from the upper/lower layer. Black rhombus depicts the moir\'e supercell. (b) Wave function densities of the first six lowest states at $\epsilon = 3.5\%$. Green rhombuses mark the supercells. (c) The moir\'e potential. (d) The light cone distribution. Black lines depict the moir\'e BZ boundaries. Red and blue/pink stars represent main and 1st/2nd Umklapp light cones. The brown dot, arrow and dashed circles represent the origin, gauge potential $\mathbf{A}$, and equi-energy rings, respectively. (e) Comparison of energy levels at $\mathbf{\Gamma}$ from volume-preserving strain (blue) and uniaxial (red dashed) strain. (f) The in-plane polarization of the first six lowest states. (g)-(i) The in-plane polarization of the first three states in case of general $\epsilon_{xx}$ and $\epsilon_{yy}$. In (f)-(i), red/blue color stands for positive/negative helicities, ellipse size stands for dipole amplitude. }
	\label{fig_realstrain}
\end{figure*}

In this section we consider moir\'e formed from uniaxial heterostrain along the zigzag direction. The strain tensor reads $S=\text{diag}(\epsilon,-\nu\epsilon)$, where the Poisson ratio $\nu=0.19$ for WSe$_2$~\cite{ccakir2014mechanical}. This geometry allows us to explore the interplay of the effects of strong geometric distortion and the pseudo-gauge potential.
Due to the different deformations along zigzag and armchair directions, the moir\'e superlattice geometry is changed significantly compared to that of a twisted or volume-preserving strained bilayer. The supercell is compressed along the short axis, forming strip-like supercells (Fig.~\ref{fig_realstrain}a).

The mini BZ is also deformed dramatically, which affects the location of Dirac cones significantly and redistributes the light cones in a very different manner (Fig.~\ref{fig_realstrain}d). A main light cone is shifted away from the origin. Besides, two of the 2nd Umklapp light cones (pink stars) stay much closer to the origin. All the light cones are close to each other in the $\mathbf{Q}_y$ direction but become well separated along $\mathbf{Q}_x$. In the presence of moir\'e potential, light cones along $\mathbf{Q}_y$ are expected to be strongly coupled. Compared with volume-preserving strain, the valley mismatch becomes smaller for uniaxial strain. This results in an overall lowering of the energy levels and band hybridization mediated by the plateau state is absent within the studied energy range since the plateau state has higher energy (Fig.~\ref{fig_realstrain}e).

As for wave function distribution, the first six states all locate around B locals (Fig.~\ref{fig_realstrain}b). Their profiles are analogous to the eigenstates of a 1D harmonic oscillator, as the potential minima are compressed to strips. Such strip-like features is a manifestation of the strong coupling between the light cones along $\mathbf{Q}_y$ direction due to the compression in the momentum space.

The light cone redistribution also affects optical properties. Since one of the main light cones is away from the origin, and there is a mixture of main and Umklapp light cones at both low and high energies near the two equi-energy circles (Fig.~\ref{fig_realstrain}d), the first six states exhibit comparable dipole strength (Fig.~\ref{fig_realstrain}f). Besides, all the eigenstates, except the ground state, are contributed by distinct light cone combinations from the case of volume-preserving strain. For the ground state that has the similar optical dipole constituent with the case of volume-preserving strain, the compressed short axis in the supercell dictates that the vertically orientated elliptical polarization becomes narrower and more linear-like (cf. Fig.~\ref{fig_strainwavefunc}b).

\subsection{General strain tuning}

\begin{figure*}
	\centering
	\includegraphics[width=1\textwidth]{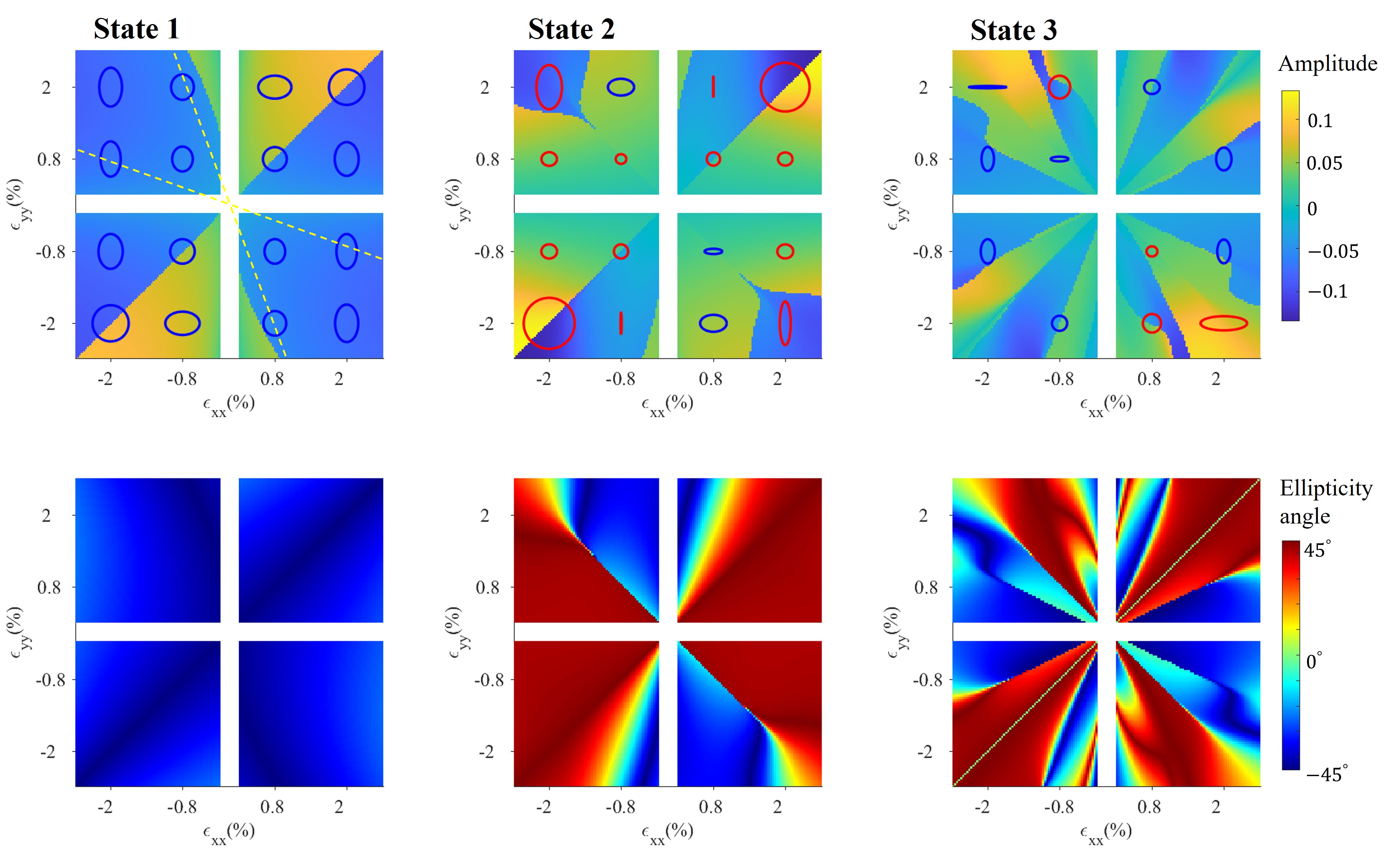}
	
	\caption{Diagrams of IXs optical properties with strain tuning. The upper three panels describe the dipole amplitude, which is measured in units of the dipoles of intralayer exciton. Positive (negative) value represents that the semi-major axis of an elliptical polarization is parallel (perpendicular) to the x axis. The two yellow dashed lines in the first panel correspond to $\epsilon_{yy}=-\nu \epsilon_{xx}$ and $\epsilon_{xx}=-\nu \epsilon_{yy}$, where $\nu = 0.19$. The lower three panels describe the ellipticity angle of polarization, where $\pm45^\circ$ stands for $\sigma_\pm$ circular polarization. Results along $\epsilon_{xx}=\epsilon_{yy}$ in the last panel are meaningless as in-plane polarization vanishes. Some specific dipole polarizations are marked as rings in the upper panel, where the center corresponds to the strain configuration, the size represents the amplitude, and red/blue color indicates positive/negative helicity.}
	\label{fig_generalstrain}
\end{figure*}

In principle, strain can be tuned independently along the two axes. To explore the effects of strain on the optical properties of moir\'e IXs more systematically, we tune the two strain components $\epsilon_{xx}$ and $\epsilon_{yy}$ continuously from volume-preserving ($\epsilon_{xx}=-\epsilon_{yy}$) to biaxial ($\epsilon_{xx}=\epsilon_{yy}$) configuration. Fig.~\ref{fig_generalstrain} shows the diagrams on the optical properties of the first three states of moir\'e IXs at $\mathbf{\Gamma}$ in the strain-parameter space. The blank areas on the diagram correspond to the situations where the superlattices become quasi-1D with gigantic periods, thus the results are not provided due to numerical limitations. The upper three panels describe the amplitude of dipoles, where positive (negative) values indicate that the semi-major axis of an elliptical polarization is parallel (perpendicular) to the x axis. The lower three panels describe the ellipticity angle of the polarization, where $\pm45^\circ$ corresponds to $\sigma_\pm$ circular polarization. The two yellow dashed lines in the first panel correspond to the cases of uniaxial strain with $\epsilon_{yy}=-\nu \epsilon_{xx}$ and $\epsilon_{xx}=-\nu \epsilon_{yy}$, where $\nu = 0.19$. The two diagonal directions in each panel describe the cases of volume-preserving and biaxial strain, respectively. In particular, a biaxially strained bilayer, whose pseudo-gauge potential vanishes, resembles a lattice-mismatched heterobilayer. Consequently, the optical properties of the former case can be employed to qualitatively understand those of the latter. Furthermore, due to similar moir\'e landscapes with identical symmetries, a biaxially strained bilayer and a twisted bilayer with equal moir\'e period share the identical in-plane polarization (cf. Fig.~\ref{fig_twistexciton}d).

%

\subsection{Mixture of twisting and heterostrain\label{section_mixture}}

In this section, we briefly discuss the mixture of twisting and heterostrain. In general, the strain and rotation operators do not commute. Also, the mixture of twisting and heterostrain will deform the BZ nontrivially, which makes the problem quite complicated. In the following, we illustrate the effects of the interplay by applying volume-preserving strain along the zigzag direction to the top layer followed by twisting.

\begin{figure}
	\centering
	\includegraphics[width=0.48\textwidth]{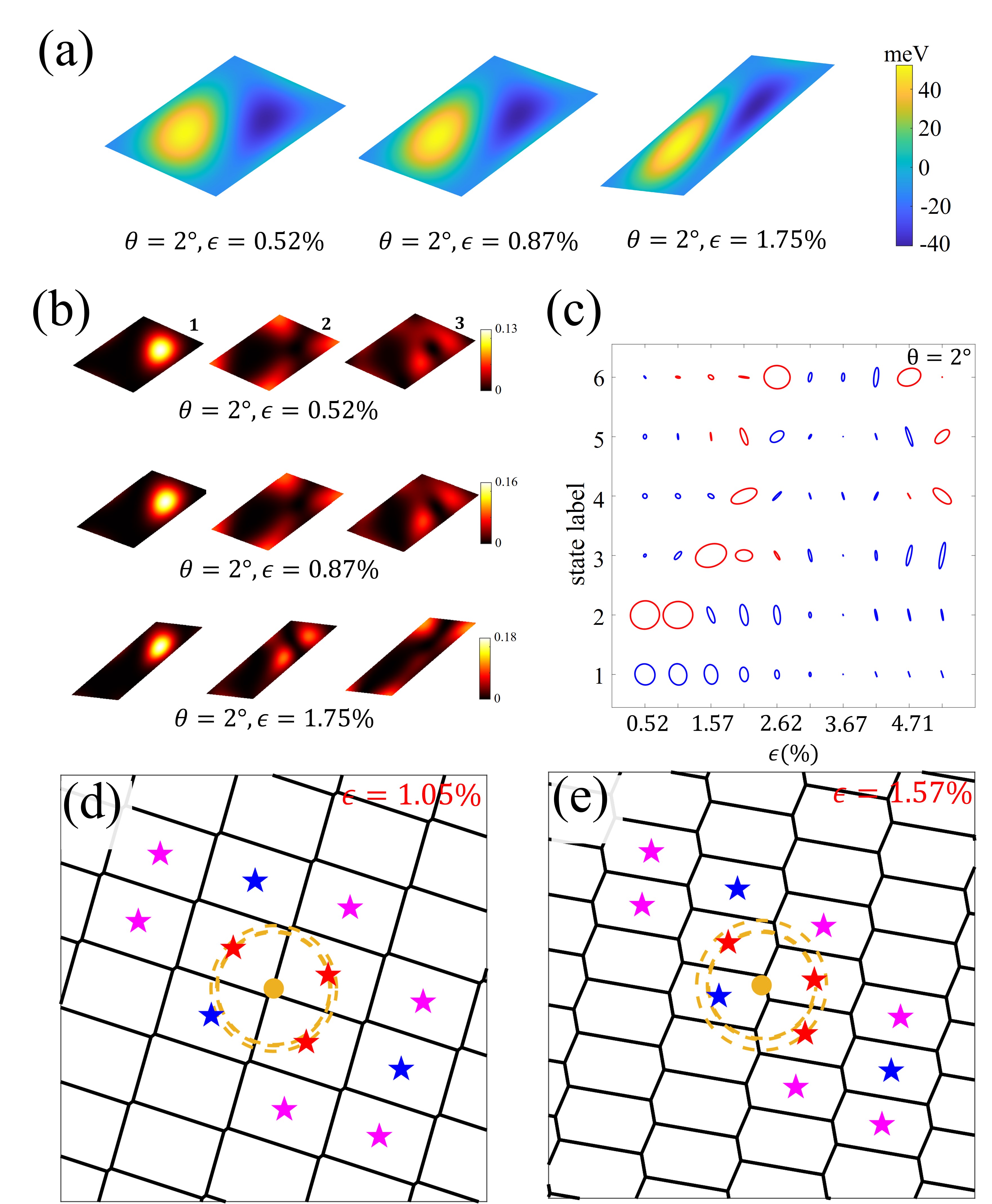}
	
	\caption{Moir\'e excitons from the mixture of twisting and volume-preserving strain. (a) Moir\'e potential landscape with fixed $\theta = 2^\circ$ while setting $\epsilon = 0.52\%$, $0.87\%$, $1.75\%$, respectively. The unit cell deforms as the strain strength approaches the size of the twist angle. (b) Wave functions of the first three states under the same conditions as (a). (c) The in-plane polarization of moir\'e excitons with fixed twist angle ($2^\circ$) and different strain intensity. Red/blue color stands for positive/negative helicities, ellipse size stands for dipole amplitude. (d) Light cone distribution at $\theta = 2^\circ$ and $\epsilon = 1.05\%$. Black lines depict the moir\'e BZ boundaries. Red and blue/pink stars represent main and 1st/2nd Umklapp light cones. The brown dot and dashed circles represent the origin and equi-energy rings, respectively. (e) Light cone distribution at $\theta = 2^\circ$ and $\epsilon = 1.57\%$.}
	\label{fig_mix}
\end{figure}

Fig.~\ref{fig_mix}a illustrates the evolution of the moir\'e potential landscape by fixing the twist angle while changing the strength of strain. When the size of strain is enlarged and approaches the twist angle, the supercell stretches and becomes a 1D strip gradually. Dirac cones in the two layers are separated in a more complicated manner in the presence of nontrivially distorted BZ and the pseudogauge potential due to the superposition of twisting and strain. This results in a redistribution of the light cones that affects the wave function distribution as well as the optical properties.

Fig.~\ref{fig_mix}b shows the wave function distribution of the three lowest states at $\mathbf{\Gamma}$ in different mixtures. Although they still exhibit localization around potential extrema at the high symmetry locals, their profiles become irregular due to the lack of both $C_3$ rotation and mirror symmetry.

When $\epsilon \leq 1.05\%$, the three main light cones maintain approximate degeneracy, so the two lowest states exhibit dominant optical dipole strength (Figs.~\ref{fig_mix}c, d). As strain is increased and approaches the size of the twist angle, main light cones are more separated and shifted toward high energy states (Fig.~\ref{fig_mix}e). Thus, high energy states can be tuned to exhibit dominant dipole strength with strain engineering (Fig.~\ref{fig_mix}c).

Strain can also be unintentionally introduced in device fabrications, which can have a general intensity and direction~\cite{shabani2021deep}. In Appendix~\ref{section_mixappendix} we provide diagrams of optical properties of moir\'e IXs in a moir\'e with a fixed twist angle but arbitrary strain intensity and direction, i.e., $S=\epsilon \begin{pmatrix}
    \text{cos}\, 2\phi & \text{sin}\, 2\phi \\
    \text{sin}\, 2\phi & -\text{cos}\, 2\phi \\
\end{pmatrix}
$, where the strain direction $\phi$ is defined by the angle between stretching direction and x axis (Figs.~\ref{fig_pt_mix_R0d5} and \ref{fig_pt_mix_R2}). Such diagrams might be utilized as a toolbox for strain estimation based on optical measurements.

\section{Summary}

To summarize, we investigate the evolution of wave function and optical properties of interlayer excitons in the moir\'e formed by different twist angles and heterostrain strength. The wave function evolution is subject to local atomic alignments and the light cone distribution. In the small twist angles or strain strength regime, densely arranged light cones dictate that wave functions are localized orbitals and optical selection rules are orbital-dependent. While in the opposite limit, sparsely arranged light cones determine that wave functions are extended Bloch-like. This results in various states possessing different rotation centers, consequently, the location-dependent optical selection rules. 

Compared with twisted moir\'e, low-energy carriers in heterostrained moir\'e are additionally affected by distorted BZ and an effective gauge potential in the momentum space, so that moir\'e excitons possess distinct wave functions and exhibit elliptically polarized optical selection rules. Due to the redistribution of light cones and 1D stripe-like wave functions in various strain configurations, high energy states can be tuned to exhibit strong optical properties. 
These results show that strain engineering can be utilized to manipulate light cone distribution and control the optical properties of moir\'e excitons.


\appendix
\section{The similarity of potentials in twisted and heterostrained moir\'e\label{section_similarity}}

\begin{figure*}
	\centering
	\includegraphics[width=1\textwidth]{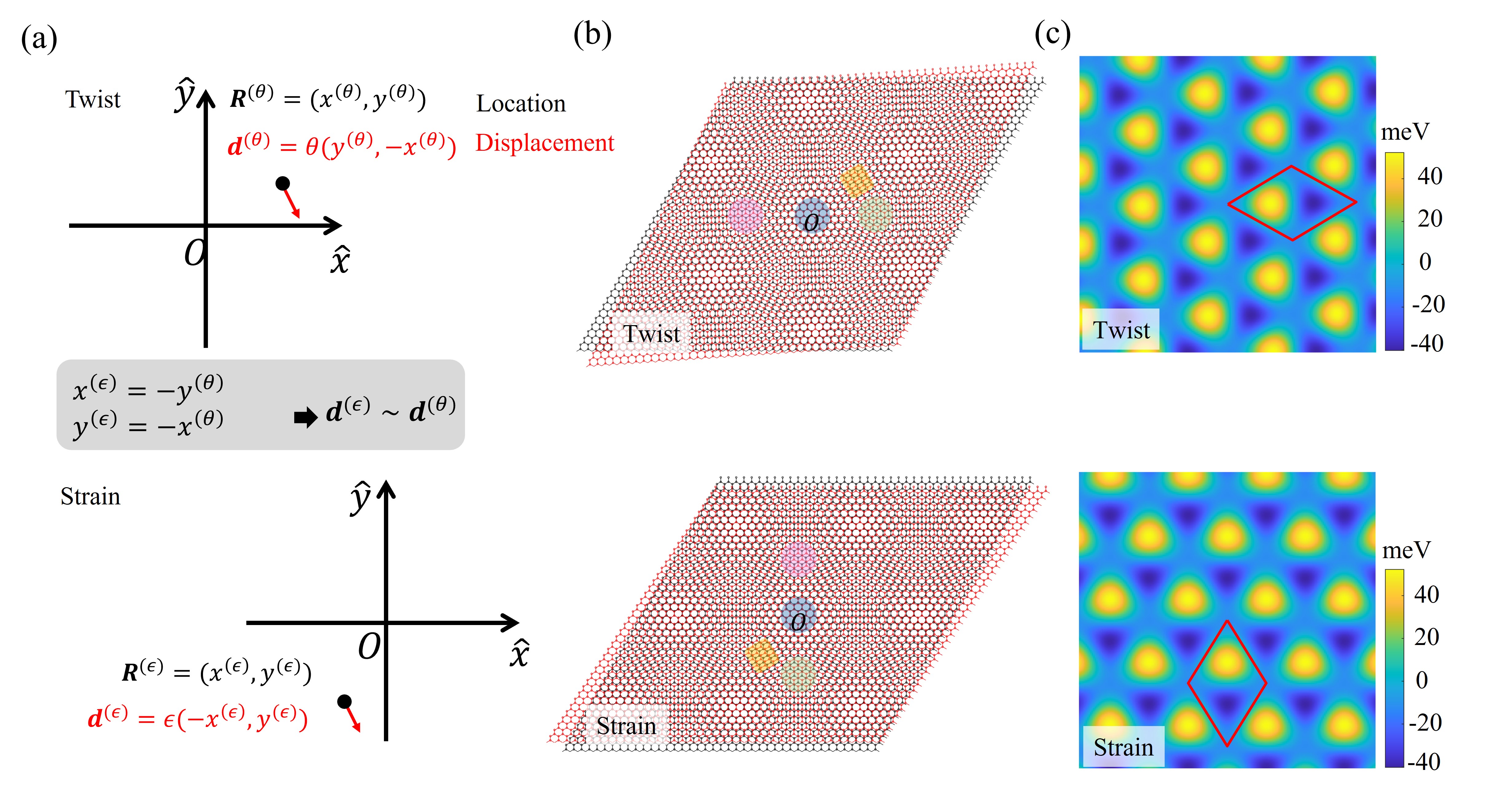}
	
	\caption{(a) Relation of the local interlayer displacements in twisted (upper panel) and volume-preserving strained (lower panel) moir\'e. A one-to-one correspondence exists between the two cases, where the local displacements (red arrows) are equivalent at two locations (black dot) in the superlattices. (b) The correspondence of the interlayer displacements results in a mapping of the local lattice-matched stacking areas in the two superlattices as represented by the shaded regions. (c) Moir\'e potentials in twisted and volume-preserving strained moir\'e. The potentials appear to be identical up to a $90^\circ$ rotation as dictated by the correspondence of the interlayer displacements.}
	\label{fig_similarity}
\end{figure*}

Our studies are focused on long-wavelength moir\'e pattern, and the similarity in the superlattice potentials in the two types of moir\'e is in the context of the low energy electron and hole (and exciton as their composite particle), which can be described well by the local approximation. In such moir\'e, the period is much larger than the lattice constant of the building block, and there exists an intermediate length scale $l$, large compared to monolayer lattice constant, but small compared to moir\'e period $a_m$. A local region of size $l$ encloses sufficient number of monolayer unit cells, whereas the interlayer atomic registry can be considered nearly uniform within this local region (as the spatial variation occurs on the scale $a_m \gg l$), where the local electronic structure can be well approximated by that of the lattice-matched bilayers with the corresponding interlayer registry.

The modelling of the moir\'e superlattices is based on such local approximations. The first step is to establish the mapping between the spatial location $\mathbf{R}$ in the moir\'e supercell and the local interlayer atomic registry $\mathbf{d}$. The mapping functions depend on the twisting between the layers, applied strain, and differences in lattice constants of the building blocks through $\mathbf{d}$. In the second step, one examines first principle calculated electronic structures of lattice-matched bilayers with various interlayer registries $\mathbf{d}$, and extracts the $\mathbf{d}$ dependence of the local properties, which, combined with the mapping $\mathbf{d}(\mathbf{R})$, will establish the moir\'e superlattice models. This approach based on the local approximation has been successfully implemented and validated in the studies of moiré superlattices.

In the following, we show explicitly the similarity between the mapping functions $\mathbf{d}(\mathbf{R})$ of the twisted moir\'e and the moir\'e from volume-preserving strain, thus similarity in their moir\'e potential landscapes. At an arbitrary location $(x,y)$, for twisted bilayer, $\mathbf{d} \approx \theta (y^{(\theta)},-x^{(\theta)})$, where the superscripts indicate the origin of the moir\'e. For heterostrained bilayer, $\mathbf{d} \approx \theta (-x^{(\epsilon)},y^{(\epsilon)})$ if volume-preserving strain is applied. One can easily identify that there is a one-to-one correspondence between locations in the two moir\'e superlattices, i.e., at $x^{(\epsilon)}=-y^{(\theta)}$ and $y^{(\epsilon)}=-x^{(\theta)}$, where the displacements become equivalent (Fig.~\ref{fig_similarity}a). Because of these correspondence in the two mapping functions $\mathbf{d}(\mathbf{R})$ (Fig.~\ref{fig_similarity}b), within the local approximation, the moir\'e potentials landscapes in the two cases differ by a $90^\circ$ rotation (Fig.~\ref{fig_similarity}c). We shall also note a subtle difference: the supercell of a twisted bilayer is a rhombus whose four sides are separated by $60^\circ$ or $120^\circ$, while the angles in a volume-preserving strained moir\'e can be slightly different, but the difference is negligible in the limit of large moir\'e period. 

In addition, the dependence of moir\'e potential on interlayer atomic registry $\mathbf{d}$ can be transformed into an equivalent form that depends on the primitive reciprocal lattice vectors of the moir\'e superlattice, i.e., $\mathbf{G}_n\cdot\mathbf{d} = \mathbf{g}_n\cdot\mathbf{R}$, where $\mathbf{G}_n$ is the primitive reciprocal lattice vectors of monolayer lattices. This format has been used in the main text (Eq.~\ref{moirepotentialformula}), where different moir\'e lattices are distinguished by the distinct reciprocal lattice vectors, 

The above local approximation also applies to moir\'e superlattices formed from other types of heterostrain or combined effects of strain and twisting. In such cases, the moir\'e realizes elongated hexagonal lattices (e.g. Fig.~\ref{fig_realstrain}), where the moir\'e potential landscapes can also be described by Eq.~(\ref{moirepotentialformula}).

\section{Optical diagram of moir\'e IXs for mixture of twisting and heterostrain\label{section_mixappendix}}

\begin{figure*}[h]
	\centering
	\includegraphics[width=1\textwidth]{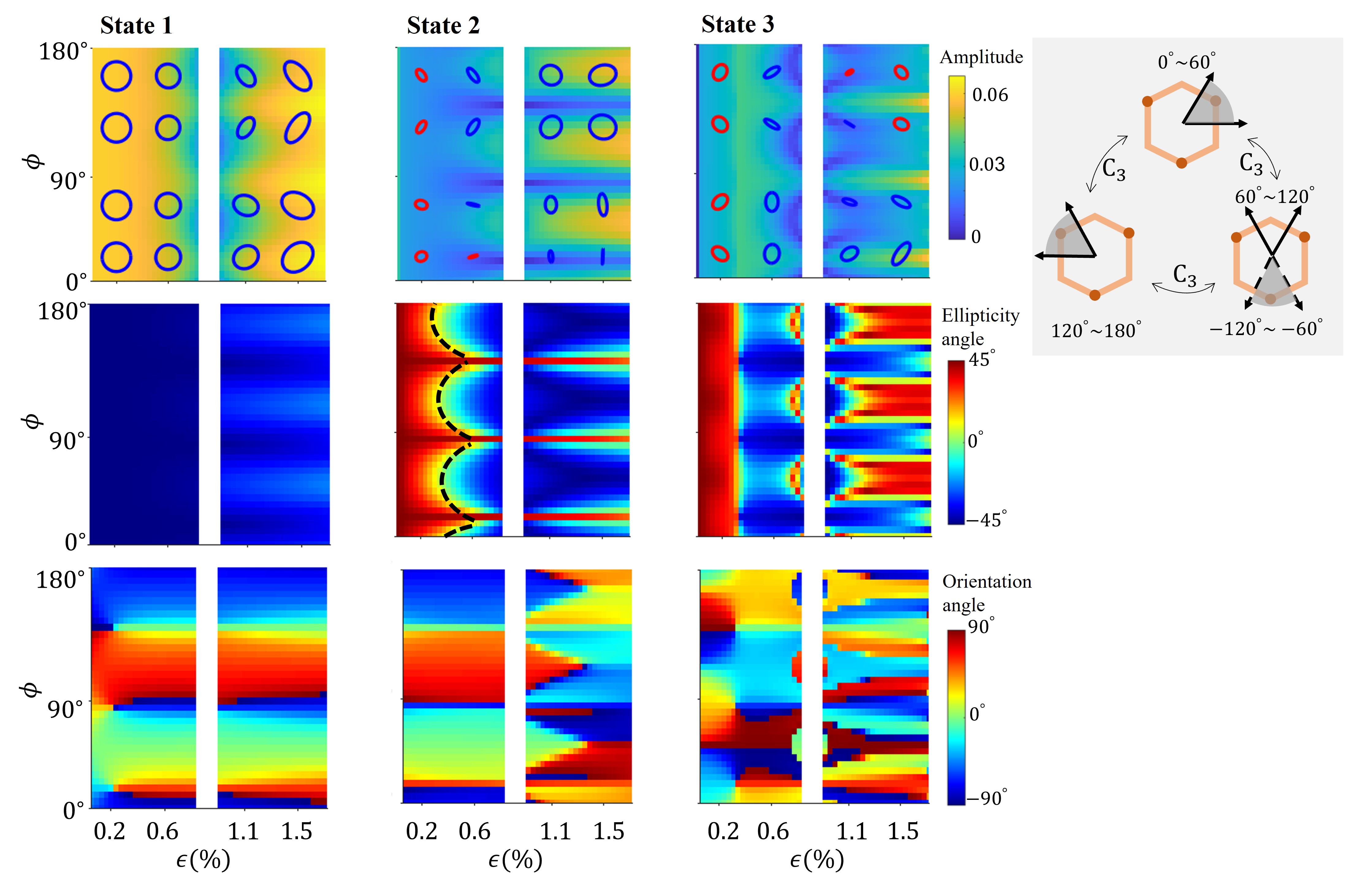}
	
	\caption{Diagrams of optical properties of IXs in the moir\'e induced by a mixture of twisting and heterostrain. Here twist angle is fixed at $0.5^\circ$, while the volume-preserving strain has its magnitude varied from $0$ to $1.75\%$ and direction from $0^\circ$ to $180^\circ$. The upper three panels describe the dipole amplitude measured in units of the dipoles of intralayer exciton. The middle three panels describe the ellipticity angle of the polarization, where $\pm45^\circ$ stands for $\sigma_\pm$ circular polarization. The lower three panels describe the orientation angle of the polarization-- angle between semi-major axis of elliptical polarization and x axis. Some specific dipole polarizations are marked in the upper panels, where the size of ellipse represents the amplitude and red/blue color indicates positive/negative helicity. The schematics on the right illustrate the equivalence of the three angular regions for applying strain, i.e., $0^\circ\sim60^\circ$, $60^\circ\sim120^\circ$, and $120^\circ\sim180^\circ$. The hexagon stands for a TMD cell and solid black arrows enclose the three angular regions.}
	\label{fig_pt_mix_R0d5}
\end{figure*}

\begin{figure*}
	\centering
	\includegraphics[width=1\textwidth]{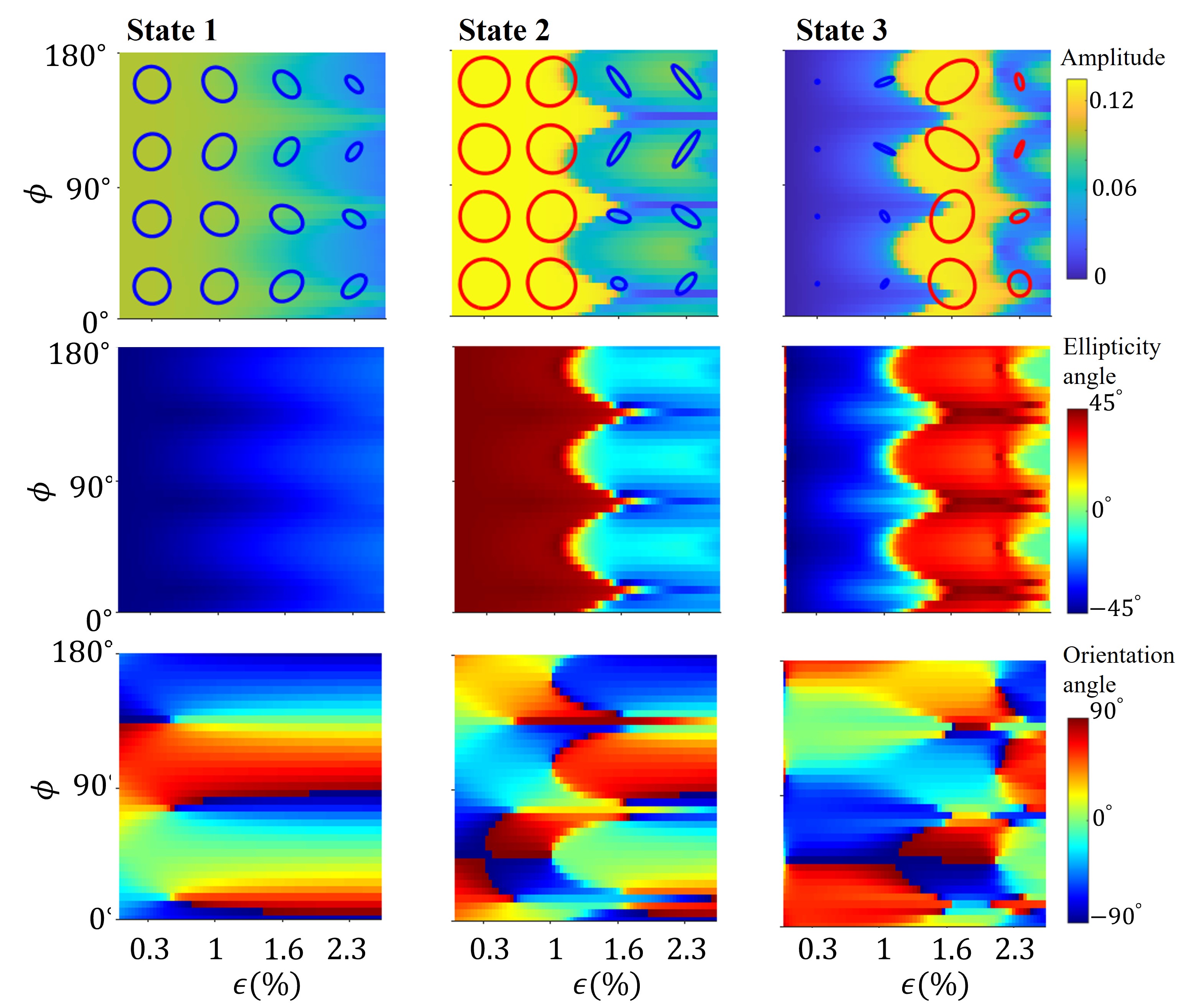}
	
	\caption{Same as Fig. \ref{fig_pt_mix_R0d5} but with twist angle equals to $2^\circ$ and volume-preserving strain whose magnitude varies from $0$ to $2.62\%$, and direction from $0^\circ$ to $180^\circ$.}
	\label{fig_pt_mix_R2}
\end{figure*}

Here we present the optical diagram of IXs in moir\'e introduced by mixture of twisting and volume-preserving heterostrain. The strain tensor reads $S=\epsilon \begin{pmatrix}
    \text{cos}\, 2\phi & \text{sin}\, 2\phi \\
    \text{sin}\, 2\phi & -\text{cos}\, 2\phi \\
\end{pmatrix}$, where the strain direction $\phi$ is defined as the angle between stretching direction and the x axis. Fig.~\ref{fig_pt_mix_R0d5} considers the small twisting angle regime ($\theta=0.5^\circ$), where strain is varied with strength $0\le\epsilon\le1.75\%$ and direction $0^\circ\le\phi\le 180^\circ$. Since the moir\'e pattern becomes quasi-1D when $\epsilon\approx\theta$, results for $0.79\%\le\epsilon\le0.96\%$ are not shown. One can see that the dipole amplitude and ellipticity (first two rows of Fig.~\ref{fig_pt_mix_R0d5}) exhibit a periodicity of $60^\circ$ with respect to $\phi$. Although not obvious, one can verify that the orientation angle (third row of Fig.~\ref{fig_pt_mix_R0d5}) modulo $60^\circ$ also shows the same periodicity. Such periodicity originates from the three-fold rotational symmetry ($C_3$) of TMDs: The crystalline structure in the region $0^\circ\sim60^\circ$ is equivalent to that in $120^\circ\sim180^\circ$ by $C_3$. While applying strain in the region $60^\circ\sim120^\circ$ is identical to that in $-120^\circ\sim-60^\circ$, which is also equivalent to $0^\circ\sim60^\circ$ by $C_3$ (see schematics in Fig.~\ref{fig_pt_mix_R0d5}). When strain is small, effects of twisting dominates, which is reflected in the circular polarization of the first state. As strain is enlarged, state exchange happens between the 2nd and the 3rd states, which is represented by the emergence of some boundaries in the diagram (e.g. black dashed curves in the middle panel). Similarly, Fig.~\ref{fig_pt_mix_R2} considers the regime of large twisting angle ($\theta=2^\circ$) with strain strength $0\le\epsilon\le2.62\%$ and direction $0^\circ\le\phi\le180^\circ$.

\begin{acknowledgments}
The authors thank Hongyi Yu for helpful discussions. The work is support by the National Key R\&D Program of China (2020YFA0309600), the Research Grant Council of Hong Kong (AoE/P-701/20), and the Croucher Senior Research Fellowship.
\end{acknowledgments}

\bibliography{references}

\end{document}